\documentclass[useAMS,usenatbib]{mn2e}

\usepackage{graphicx}
\usepackage{amsmath}
\usepackage{amssymb}

\newcommand{\ion}[2]{\mbox{#1$\,$\textsc{#2}}}

\title[Gas and dark matter in the Sculptor group: NGC~300]{Gas and dark matter in the Sculptor group: NGC~300}
\author[T.~Westmeier, R.~Braun, and B.~S.~Koribalski]{T.~Westmeier$^{1}$\thanks{E-mail:
tobias.westmeier@csiro.au}, R.~Braun$^{1}$, and B.~S.~Koribalski$^{1}$\\
$^{1}$Australia Telescope National Facility, CSIRO Astronomy and Space Science, PO Box 76, Epping NSW 1710, Australia}

\begin{document}
  \date{Accepted 1988 December 15. Received 1988 December 14; in original form 1988 October 11}
  \pagerange{\pageref{firstpage}--\pageref{lastpage}} \pubyear{2009}
  \maketitle
  \label{firstpage}
  
  \begin{abstract}
    We used the Australia Telescope Compact Array to map a large field of approximately $2^{\circ} \times 2^{\circ}$ around the Sculptor group galaxy NGC~300 in the 21-cm line emission of neutral hydrogen. We achieved a $5 \sigma$ \ion{H}{i} column density sensitivity of $10^{19}~\mathrm{cm}^{-2}$ over a spectral channel width of $8~\mathrm{km \, s}^{-1}$ for emission filling the $180'' \times 88''$ synthesised beam. The corresponding \ion{H}{i} mass sensitivity is $1.2 \times 10^{5}~\mathrm{M}_{\odot}$, assuming a distance of $1.9~\mathrm{Mpc}$. For the first time, the vast \ion{H}{i} disc of NGC~300 has been mapped over its entire extent at a moderately high spatial resolution of about $1~\mathrm{kpc}$.
    
    NGC~300 is characterised by a dense inner \ion{H}{i} disc, well aligned with the optical disc of $290^{\circ}$ orientation angle, and an extended outer \ion{H}{i} disc with a major axis of more than $1^{\circ}$ on the sky (equivalent to a diameter of about $35~\mathrm{kpc}$) and a different orientation angle of $332^{\circ}$. A significant fraction (about 43~per cent) of the total detected \ion{H}{i} mass of $1.5 \times 10^{9}~\mathrm{M}_{\odot}$ resides within the extended outer disc. We fitted a tilted ring model to the velocity field of NGC~300 to derive the rotation curve out to a radius of $18.4~\mathrm{kpc}$, almost twice the range of previous rotation curve studies. The rotation curve rises to a maximum velocity of almost $100~\mathrm{km \, s}^{-1}$ and then gently decreases again in the outer disc beyond a radius of about $10~\mathrm{kpc}$. Mass models fitted to the derived rotation curve yield good fits for Burkert and NFW dark matter halo models, whereas pseudo-isothermal halo models and MOND-based models both struggle to cope with the declining rotation curve.
    
    We also observe significant asymmetries in the outer \ion{H}{i} disc of NGC~300, in particular near the edge of the disc, which are possibly due to ram pressure stripping of gas by the intergalactic medium (IGM) of the Sculptor group. Our estimates show that ram pressure stripping can occur under reasonable assumptions on the density of the IGM and the relative velocity of NGC~300. The asymmetries in the gas disc suggest a proper motion of NGC~300 toward the south-east. At the same time, our data exclude IGM densities of significantly higher than $10^{-5}~\mathrm{cm}^{-3}$ in the vicinity of NGC~300, as otherwise the outer gas disc would have been stripped.
  \end{abstract}
  
  
  \section{Introduction}
  
  At distances in the range of about $2$ to $5~\mathrm{Mpc}$, the Sculptor group is among the nearest galaxy groups beyond the Local Group \citep{Jerjen1998}. It forms an elongated filament of galaxies and comprises a number of separate subgroups at different distances along the line of sight. At a distance of about $2~\mathrm{Mpc}$, the nearest of these subgroups consists of NGC~55, NGC~300, and possibly two or more known dwarf spheroidal companions (\citealt{Karachentsev2003}; Koribalski et al., in prep.). NGC~55 and NGC~300 are medium-sized spiral galaxies of type SB(s)m and SA(s)d, respectively \citep{deVaucouleurs1991}. Their proximity makes them preferential targets for deep \ion{H}{i} observations with high spatial resolution and \ion{H}{i} mass sensitivity.
  
  Some of the basic properties and physical parameters of NGC~300, including the results of this work, are listed in Table~\ref{tab_ngc300}. The distance towards NGC~300 has been measured with great accuracy and through different methods. \citet{Rizzi2006} used the tip of the red giant branch to determine distance moduli of $(m - M)_{0} = 26.30 \pm 0.03 \pm 0.12~\mathrm{mag}$ and $26.36 \pm 0.02 \pm 0.12~\mathrm{mag}$ through two different statistical methods as part of the Araucaria Project \citep{Gieren2005b}. Their results are consistent with the distance moduli of $26.43 \pm 0.04 \pm 0.05$ and $26.37 \pm 0.05 \pm 0.03~\mathrm{mag}$ derived from Cepheid variables by \citet{Gieren2004} and \citet{Gieren2005a}, respectively. Based on these measurements we therefore adopt a distance of $1.9~\mathrm{Mpc}$ for NGC~300 throughout this paper.
  
  Early \ion{H}{i} imaging of NGC~300 was carried out by \citet{Shobbrook1967} using the 64-m Parkes radio telescope. They determined an \ion{H}{i} mass of $2.1 \times 10^{9}~\mathrm{M}_{\odot}$ and a total mass of $(2.5 \pm 0.5) \times 10^{10}~\mathrm{M}_{\odot}$ from the rotation curve under the assumption of a distance of $1.9~\mathrm{Mpc}$. They also noticed that the velocity field of NGC~300 appears strongly distorted which they attributed to the gravitational influence of a large \ion{H}{i} cloud to the south-east of the galaxy. This cloud, however, was later found to be most likely part of the Magellanic Stream \citep*{Mathewson1975,Rogstad1979}.
  
  \begin{table}
    \centering
    \caption{Properties of NGC~300. The separation between inner and outer \ion{H}{i} disc was made at a column density level of $5 \times 10^{20}~\mathrm{cm}^{-2}$. Note that all errors are statistical uncertainties and do not reflect possible systematic errors. References: [1]~\citet{deVaucouleurs1991}; [2]~\citet{Gieren2005a}; [3]~\citet{Rizzi2006}; [4]~\citet{Carignan1985}. Values without reference are the result of this study.}
    \label{tab_ngc300}
    \begin{tabular}{lrll}
      \hline
      Parameter & Value & Unit & Ref. \\
      \hline
      Type                   &                             SA(s)d &                         & [1]     \\
      $\alpha$ (J2000)       &   $00^{\rm h} 54^{\rm m} 53\fs{}4$ &                         &         \\
      $\delta$ (J2000)       &    ${-37}^{\circ} 41' 02\farcs{}6$ &                         &         \\
      Distance               &                              $1.9$ &          $\mathrm{Mpc}$ & [2] [3] \\
      $R_{25}$               &                       $9\farcm{}8$ &                         & [4]     \\
      Radial velocity        &                                    &                         &         \\
      \quad barycentric      &                        $144 \pm 2$ & $\mathrm{km \, s}^{-1}$ &         \\
      \quad LSR              &                        $136 \pm 2$ & $\mathrm{km \, s}^{-1}$ &         \\
      Position angle$^1$     &                                    &                         &         \\
      \quad inner disc       &          $290\fdg{}0 \pm 0\fdg{}5$ &                         &         \\
      \quad outer disc       &          $331\fdg{}7 \pm 0\fdg{}3$ &                         &         \\
      Inclination            &     $40^{\circ} \ldots 50^{\circ}$ &                         &         \\
      Max. rot. velocity     &                     $98.8 \pm 3.1$ & $\mathrm{km \, s}^{-1}$ &         \\
      Integrated flux        &               $1.72 \times 10^{3}$ & $\mathrm{Jy \, km \, s}^{-1}$ &   \\
      \ion{H}{i} mass        &                $1.5 \times 10^{9}$ &             $\mathrm{M}_{\odot}$ &         \\
      \quad inner disc       &                $8.5 \times 10^{8}$ &             $\mathrm{M}_{\odot}$ &         \\
      \quad outer disc       &                $6.5 \times 10^{8}$ &             $\mathrm{M}_{\odot}$ &         \\
      Gas mass$^2$           &      $(1.9 \pm 0.2) \times 10^{9}$ &             $\mathrm{M}_{\odot}$ &         \\
      Stellar mass$^2$       &      $(1.0 \pm 0.1) \times 10^{9}$ &             $\mathrm{M}_{\odot}$ &         \\
      Total mass$^3$         &     $(2.9 \pm 0.2) \times 10^{10}$ &             $\mathrm{M}_{\odot}$ &         \\
      \hline
      \multicolumn{4}{l}{\footnotesize $^1$~w.r.t.\ the J2000.0 equatorial coordinate system} \\
      \multicolumn{4}{l}{\footnotesize $^2$~across our tilted ring area ($0.9 < r < 18.4~\mathrm{kpc}$)} \\
      \multicolumn{4}{l}{\footnotesize $^3$~within $r = 18.4~\mathrm{kpc}$}
    \end{tabular}
  \end{table}
  
  A larger area around NGC~300 was mapped in \ion{H}{i} by \citet{Mathewson1975} with the 64-m Parkes radio telescope. Within their brightness temperature sensitivity of $0.3~\mathrm{K}$ over $4.1~\mathrm{km \, s}^{-1}$ and across the $15~\mathrm{arcmin}$ Parkes beam they detected several extended gas clouds with velocities comparable to those found across NGC~300, including a long `\ion{H}{i} tail' extending about $2^{\circ}$ to the southeast of the galaxy. While \citet{Mathewson1975} assumed that these clouds were associated with NGC~300, there has been a long debate over the past decades whether they might instead be fragments of the Magellanic Stream that runs across the same part of the sky \citep[e.g.][]{Haynes1979,Putman2003}.
  
  A more detailed \ion{H}{i} image of NGC~300 was obtained by \citet{Rogstad1979} with the twin-element interferometer of the Owens Valley Radio Observatory. At a much higher angular resolution of $2' \times 3'$ half-power beam width they were able to derive a detailed radial velocity map of the inner part of the galaxy. By fitting a tilted ring model to the velocity field they were able to describe the distorted velocity contours of NGC~300 as a result of varying inclination and position angle across the \ion{H}{i} disc. The rotation curve obtained by \citet{Rogstad1979} from the tilted ring model is basically flat beyond a radius of about $10~\mathrm{arcmin}$ with a turnover velocity of $94~\mathrm{km \, s}^{-1}$ at $16~\mathrm{arcmin}$ radius. They suggest that the warping of the \ion{H}{i} disc of NGC~300 could have been caused by a close encounter with another galaxy or massive \ion{H}{i} cloud within the past $10^{9}$~years. Similar \ion{H}{i} synthesis observations of NGC~300 with somewhat higher angular resolution ($50'' \times 50''$ half-power beam width) were obtained by \citet*{Puche1990} who employed a mosaic of five pointings with the Very Large Array (VLA).
  
  Unfortunately, the synthesis images obtained by \citet{Rogstad1979} and \citet{Puche1990} did not reveal the extent and structure of the outer disc of NGC~300 due to insufficient field of view and lack of sensitivity. We therefore decided to obtain deep \ion{H}{i} observations of a large field of $2^{\circ} \times 2^{\circ}$ around NGC~300 in a mosaic of 32~pointings with the Australia Telescope Compact Array (ATCA). The aim of these observations was to map the entire extent of the \ion{H}{i} disc of NGC~300, determine its structure, and dynamics, and search for extra-planar gas in the vicinity of the galaxy. The observations and results of this project are described in this paper. Similar observations of the Sculptor group galaxy NGC~55 will be presented in a separate paper.
  
  This paper is organized as follows: Section~\ref{sect_observations} describes our observations and data reduction procedure. In Section~\ref{sect_results} we discuss our results, including the general physical parameters of NGC~300 and the results of the rotation curve analysis. In Section~\ref{sect_massmodels} we describe the fitting of mass models, with and without dark matter, to the observed rotation curve. In Section~\ref{sect_discussion} we discuss the possible origin of the extended outer gas disc of NGC~300 and present evidence of the distortion of the gas disc by ram pressure. Finally, Section~\ref{sect_summary} gives a summary of our results and conclusions.
  
  \section{Observations and data reduction}
  \label{sect_observations}
  
  The observations of NGC~300 were carried out in late 2007 and 2008 with the six 22-m antennas of the Australia Telescope Compact Array. Five of the antennas are moveable along an east-west track with a total length of $3~\mathrm{km}$, whereas the sixth antenna is fixed and separated by another $3~\mathrm{km}$ from the western end of the track.
  
  The total observing time of $96~\mathrm{h}$ was equally divided among the EW352 and EW367 array configurations to improve baseline coverage. Because of the large spatial separation between antenna~6 and all other antennas in these configurations we only used data from antennas~1 to~5 in our analysis. The resulting minimum and maximum baselines were $31$ and $367~\mathrm{m}$, respectively.
  
  NGC~300 was simultaneously observed in the 21-cm line emission of neutral hydrogen \citep{Wild1952} and in 21-cm radio continuum emission, using the two independent frequency chains of the ATCA. For the \ion{H}{i} observations we used a bandwidth of $8~\mathrm{MHz}$ centred at a frequency of $1420~\mathrm{MHz}$. The resulting velocity coverage is about $1600~\mathrm{km \, s}^{-1}$ with a channel separation of $3.3~\mathrm{km \, s}^{-1}$ and an effective velocity resolution of about $4~\mathrm{km \, s}^{-1}$. The radio continuum data were obtained over a bandwidth of $128~\mathrm{MHz}$ centred at $1384~\mathrm{MHz}$. Only the \ion{H}{i} data will be discussed in this paper.
  
\begin{figure}
    \centering
    \includegraphics[width=\linewidth]{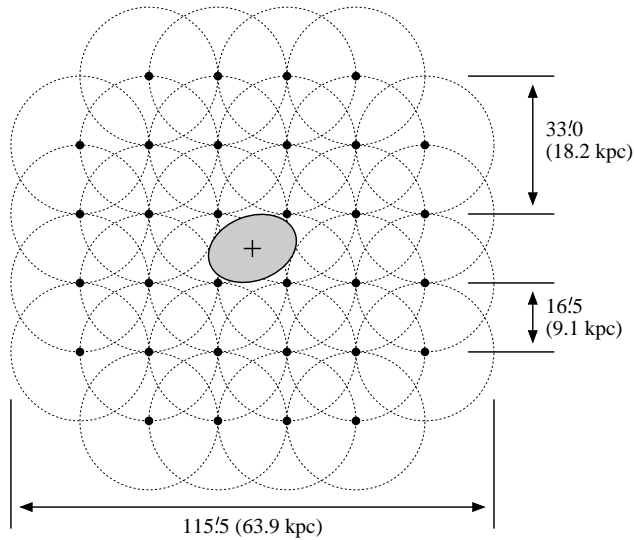}
    \caption{Outline of our ATCA mosaic of 32~pointings around NGC~300 (north is up). Pointing centres are marked by black points with the dotted circles indicating the half-power beam width of $33~\mathrm{arcmin}$ of the ATCA antennas at $\lambda = 21~\mathrm{cm}$. The black cross marks the central position of the entire mosaic (and of NGC~300) at $\alpha = 00^{\rm h}54^{\rm m}53\fs{}5$ and $\delta = {-37}^{\circ}41'04''$ in J2000 equatorial coordinates. The grey ellipse outlines the approximate size and orientation of the optical disc of NGC~300.}
    \label{fig_mosaic}
  \end{figure}

  In order to map a large field of about $2^{\circ} \times 2^{\circ}$ (corresponding to a projected size of about $65 \times 65~\mathrm{kpc}^{2}$) we covered NGC~300 with a large mosaic of 32 separate pointings. Adjacent pointings were separated by $16\farcm{}5$ which corresponds to approximately half the half-power beam width of the ATCA antennas at a wavelength of $\lambda = 21~\mathrm{cm}$. The geometry of the entire mosaic is outlined in Fig.~\ref{fig_mosaic}. The integration time per pointing in each array configuration was $1.5~\mathrm{h}$. The resulting spectral rms noise level at $3.3~\mathrm{km \, s}^{-1}$ channel separation is about $5.5~\mathrm{mJy}$ per beam which is equal to the theoretical noise level.
  
  At the beginning of each observing run we spent about 15 to 20~minutes on the ATCA's standard flux calibration source, PKS~1934$-$638, which was used for bandpass and flux calibration of the data. Every 45~minutes during our observing runs we integrated for 5~minutes on our gain amplitude and phase calibrator, PKS~0008$-$421, which has a $20~\mathrm{cm}$ flux of about $4.5~\mathrm{Jy}$ and a separation from NGC~300 of about $9\fdg{}5$. PKS~0008$-$421 was used to calibrate gain amplitudes and phases as a function of time.
  
  \begin{figure*}
    \centering
    \includegraphics[width=\linewidth]{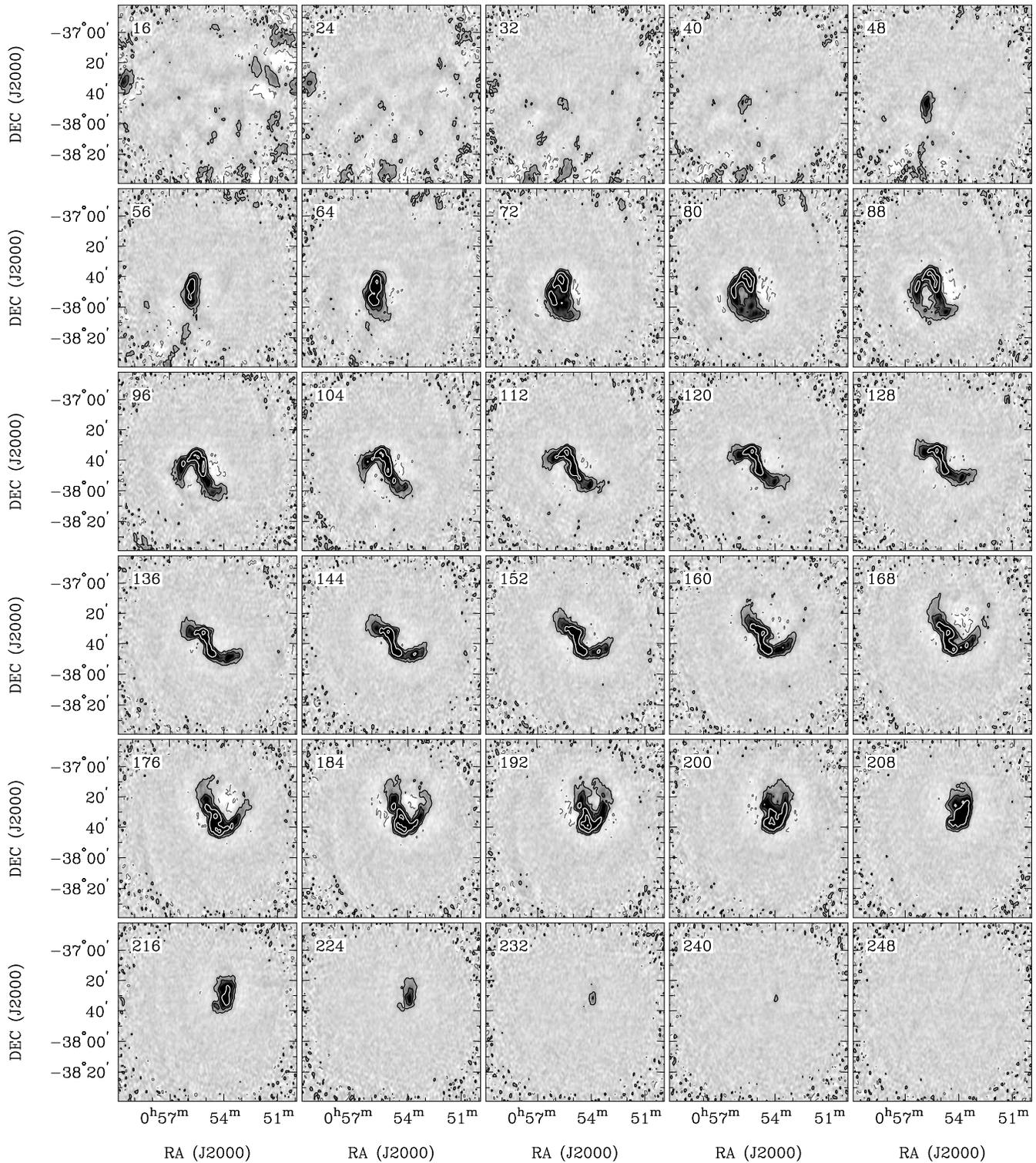}
    \caption{Channel maps of the ATCA \ion{H}{i} data cube of NGC~300 in the range of $v_{\rm LSR} = 16$ to $248~\mathrm{km \, s}^{-1}$ (corresponding to barycentric velocities of $24$ to $256~\mathrm{km \, s}^{-1}$). The greyscale images range from $-20$ to $+100~\mathrm{mJy}$ per beam. The contour levels are $-15$ (dashed), $15$, $50$, $150$, and $500~\mathrm{mJy}$ per beam. The increased noise along the edges of the maps is the result of primary beam attenuation. Channels containing Galactic \ion{H}{i} emission were not deconvolved.}
    \label{fig_chanmaps}
  \end{figure*}
  
  \begin{figure*}
    \centering
    \includegraphics[width=0.88\linewidth]{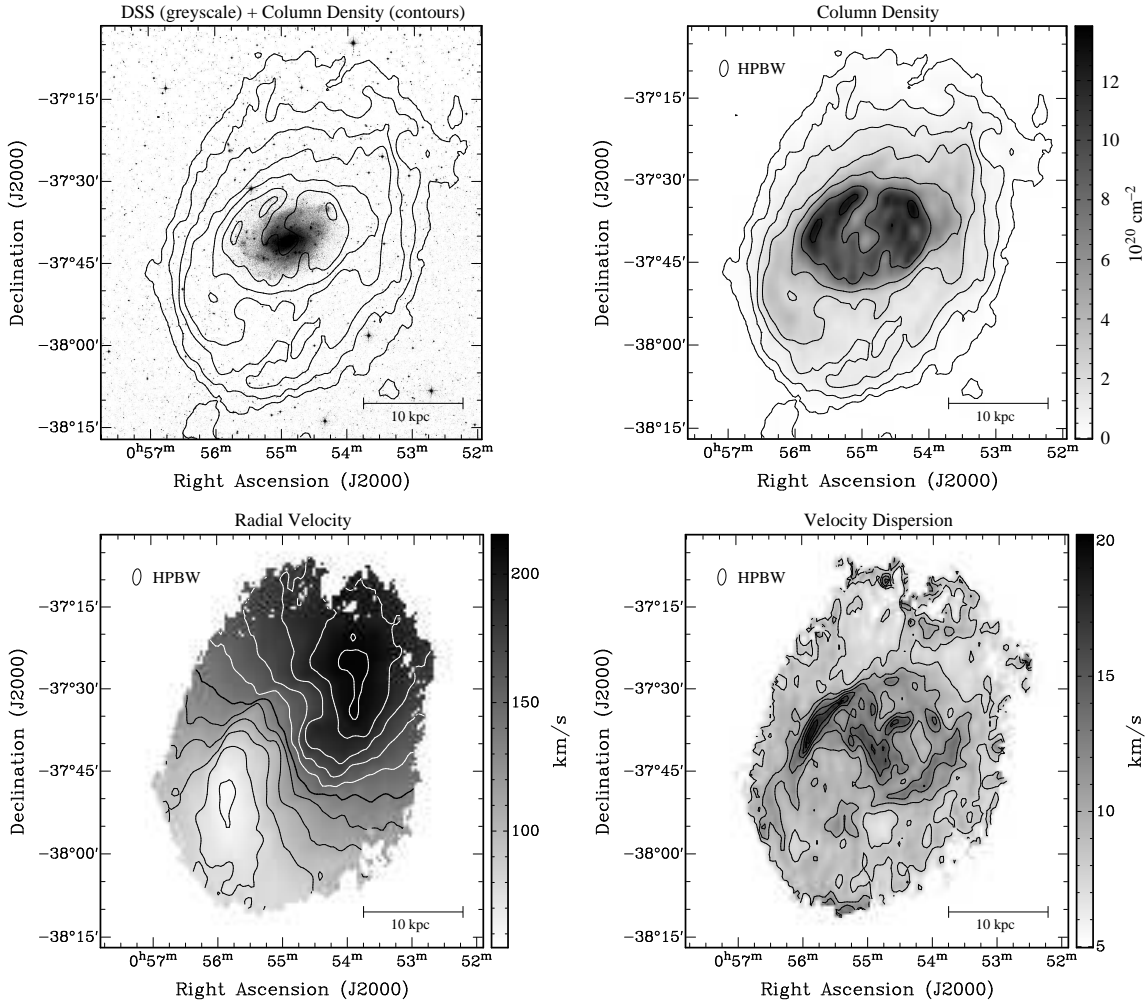}
    \caption{Upper-left panel: Optical image of NGC~300 from the Digitized Sky Survey (in logarithmic scaling) with \ion{H}{i} column density contours overlaid. Upper-right panel: \ion{H}{i} column density map of NGC~300 derived from the zeroth moment under the assumption that the gas is optically thin. The black \ion{H}{i} column density contours in this and the previous map correspond to $0.1$, $0.4$, $1$, $2$, $4$, $8$, and $12 \times 10^{20}~\mathrm{cm}^{-2}$. Lower-left panel: LSR radial velocity field derived from Gauss--Hermite polynomials. The contours are drawn at intervals of $15~\mathrm{km \, s}^{-1}$ centred on the systemic velocity of NGC~300 of $v_{\rm LSR} = 136~\mathrm{km \, s}^{-1}$ (bold contour line; equivalent to $144~\mathrm{km \, s}^{-1}$ in the barycentric reference frame). Lower-right panel: Velocity dispersion map derived from Gauss--Hermite polynomials. The contours correspond to $8$, $10$, $12$, $14$, $16$, $18$, and $20~\mathrm{km \, s}^{-1}$.}
    \label{fig_coldens}
  \end{figure*}
  
  The data were reduced and analysed with the Astronomical Image Processing System (AIPS; \citealt{Greisen1990}), the Multichannel Image Reconstruction, Image Analysis and Display (\textsc{Miriad}) software package \citep{Sault1995}, and the Groningen Image Processing System (GIPSY; \citealt{Allen1985,vanderHulst1992}) in a combined approach. We first used the task \textsc{atlod} to read the raw data into AIPS for the purpose of flagging bad pixels in the time-baseline domain using the interactive task \textsc{spflg}. Next, we imported the flagged data into \textsc{Miriad}. The flagging table produced by AIPS was then applied via the \textsc{Miriad} task \textsc{fgflag} before proceeding with some additional flagging of shadowed antennas, \ion{H}{i} absorption in the primary calibrator, PKS~1934$-$638, and otherwise corrupted visibilities. We then carried out the standard data reduction procedure, making use of \textsc{Miriad}'s dedicated mosaicking capabilities which include automatic corrections for primary beam attenuation.
  
  For the image cube and beam we applied a robust weighting of visibilities with a robustness parameter of zero. The resulting synthesised beam is elliptical with a full width at half maximum (FWHM) of $180'' \times 88''$ and a position angle of ${-6}^{\circ}$ with respect to north. Furthermore, we employed a pixel size of $30~\mathrm{arcsec}$ in the image domain and a resolution and channel separation of $8~\mathrm{km \, s}^{-1}$ through Hanning smoothing in the spectral domain. The resulting spectral rms noise level is $3.5~\mathrm{mJy}$ per beam, corresponding to a $5 \sigma$ \ion{H}{i} column density sensitivity of $1.0 \times 10^{19}~\mathrm{cm}^{-1}$ per spectral channel. This translates into a $5 \sigma$ \ion{H}{i} mass sensitivity of $1.2 \times 10^{5}~\mathrm{M}_{\odot}$ at a distance of $1.9~\mathrm{Mpc}$.
  
  \begin{figure*}
    \centering
    \includegraphics[width=0.87\linewidth]{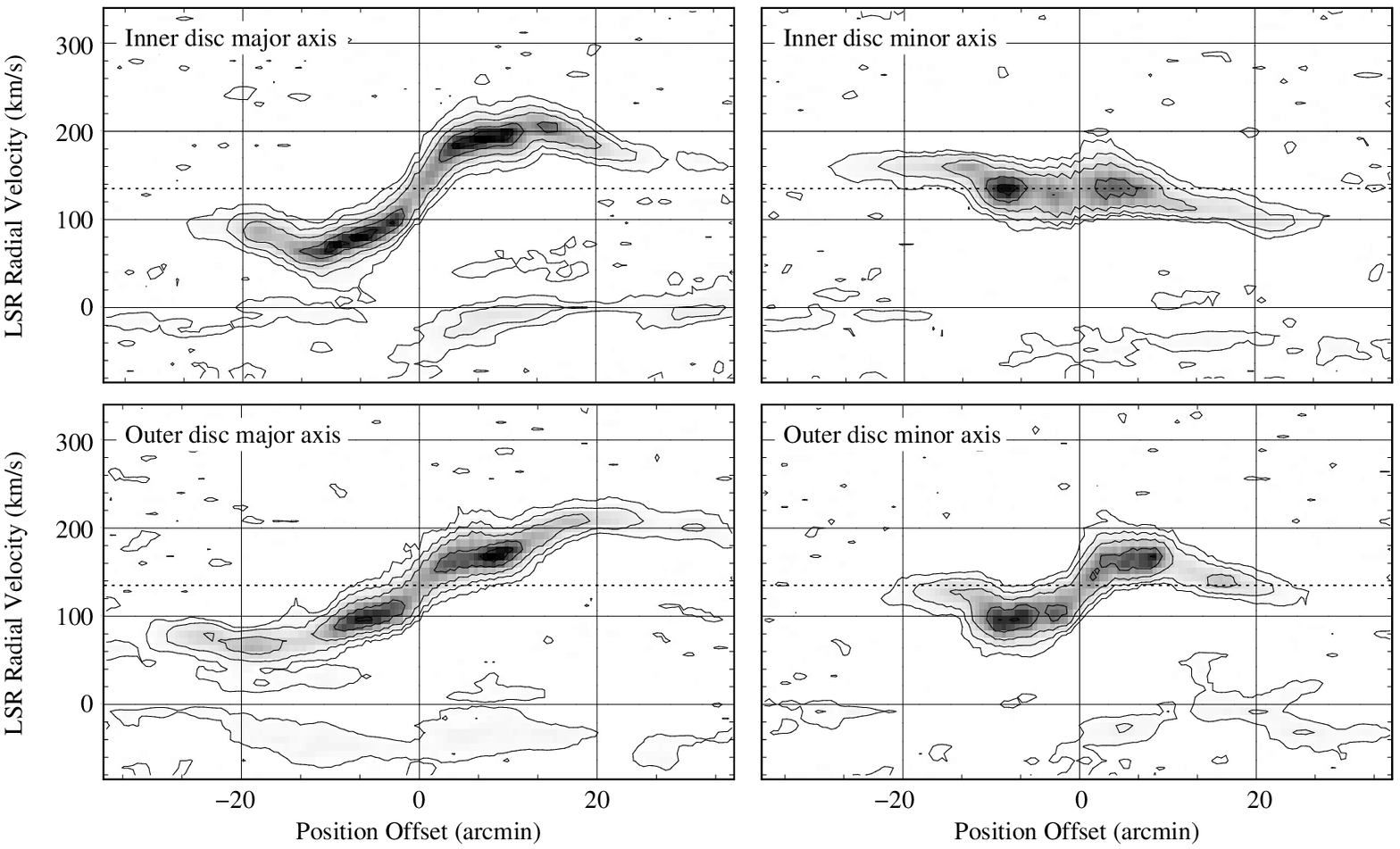}
    \caption{Position-velocity diagrams along the major and minor axis of the inner disc (top panels, with position angles of $\varphi = 290^{\circ}$ and $200^{\circ}$, respectively) and along the major and minor axis of the outer disc (bottom panels, with $\varphi = 332^{\circ}$ and $242^{\circ}$, respectively). Contour levels are $5$, $30$, $100$, $300$, and $500~\mathrm{mJy}$ per beam. The dotted line indicates the systemic velocity of NGC~300 of $v_{\rm sys} = 136~\mathrm{km \, s}^{-1}$ in the LSR frame (equivalent to $144~\mathrm{km \, s}^{-1}$ in the barycentric reference frame). The offset of $0$ corresponds to the centre of NGC~300. Emission near $v_{\rm LSR} = 0~\mathrm{km \, s}^{-1}$ is due to Galactic foreground.}
    \label{fig_posivelo}
  \end{figure*}
  
  To reduce the impact of beam sidelobes caused by our incomplete aperture synthesis we subsequently deconvolved the image data cube using the CLEAN algorithm originally developed by \citet{Hoegbom1974}. We actually applied the Steer CLEAN algorithm \citep{Steer1984} implemented in the \textsc{Miriad} task \textsc{mossdi} with a flux density cutoff of $10~\mathrm{mJy}$ per beam equivalent to about three times the rms noise level at $8~\mathrm{km \, s}^{-1}$ channel width. Individual channel maps of the final data cube are shown in Fig.~\ref{fig_chanmaps}.
  
  The zeroth, first, and second moment maps of NGC~300 were calculated using the AIPS task \textsc{momnt}. For this purpose, we first multiplied the data cube with the theoretical gain across the mosaic to obtain a data cube with a constant noise level at all positions and frequencies. Next, we created a mask from the data cube by Hanning-smoothing the frequency axis over three channels, convolving the spatial dimension with a Gaussian of 5~pixels FWHM, and then masking all pixels in the smoothed cube that had a flux density value of less than $4.5~\mathrm{mJy}$ per beam. In \textsc{momnt}, only those pixels of the original data cube that had not been masked in the smoothed cube were used in calculating the moment maps. Finally, the moment maps were divided again by the gain model to recover the original gain levels across the mosaic.
  
  There are several ways of determining the mean radial velocity of an \ion{H}{i} line, including the calculation of the first moment of the spectrum, the fitting of one or more Gaussian functions to the spectral line, or simply by using the velocity of the peak of the \ion{H}{i} line. The impact of these and other methods on the outcome of rotation curve fits was studied in detail by \citet{deBlok2008} based on a sample of galaxies observed as part of The \ion{H}{i} Nearby Galaxy Survey (THINGS; \citealt{Walter2008}). They found the most suitable method to be the fitting of Gauss--Hermite polynomials \citep{vanderMarel1993} with an additional skewness parameter, $h_{3}$, to the spectral lines. The fitting function used by \citet{deBlok2008} and implemented in the GIPSY task \textsc{xgaufit} is a modified Gauss--Hermite polynomial including only the zeroth- and third-order terms, namely
  \begin{equation}
    A(y) = a \exp \! \left( -y^{2} / 2 \right) \left( 1 + h_{3} \frac{2 y^{3} - 3 y}{\sqrt{3}} \right)
  \end{equation}
  with the substitution $y = (x - b) / c$. The free parameters of the fit are $a$, $b$, $c$, and the aforementioned skewness parameter, $h_{3}$. The advantage of using Gauss--Hermite polynomials instead of pure Gaussian functions is that the former will account for asymmetries in the line profile that can be characterised by $h_{3}$. In our analysis we used the position of the maximum of $A(y)$ to derive the radial velocity field.
  
  \section{Results}
  \label{sect_results}
  
  \subsection{General parameters}
  
  The total \ion{H}{i} column density map of NGC~300, derived from the zeroth moment of the data cube, is shown in Fig.~\ref{fig_coldens}.\footnote{Note that the assumption was made that the \ion{H}{i} gas is optically thin. In the innermost regions of NGC~300 this assumption may not be valid, and the true column densities may be higher.} The \ion{H}{i} disc appears to be significantly more extended than the optical disc and shows a clear separation into an inner and outer disc. The inner disc is characterised by high column densities of $N_{\rm HI} \gtrsim 5 \times 10^{20}~\mathrm{cm}^{-2}$ and appears to be aligned with the stellar disc. In contrast, the outer disc has lower column densities of typically $N_{\rm HI} \lesssim 2 \times 10^{20}~\mathrm{cm}^{-2}$ and is spatially much more extended. Furthermore, the transition between the inner and outer disc seems to go along with a systematic twist resulting in very different orientation angles of the two discs. This twist is also evident in the radial velocity map in Fig.~\ref{fig_coldens} and the position-velocity diagrams in Fig.~\ref{fig_posivelo}.
  
  Owing to our large mapping area and sufficiently high sensitivity, this is the first time that the entire extent (down to a column density level of about $10^{19}~\mathrm{cm}^{-2}$) of the \ion{H}{i} disc of NGC~300 has been revealed in such detail. Within the $10^{19}~\mathrm{cm}^{-2}$ column density level, the major axis of the \ion{H}{i} disc of NGC~300 spans slightly more than $1^{\circ}$ on the sky which is equivalent to a linear diameter of about $35~\mathrm{kpc}$. Earlier \ion{H}{i} observations of NGC~300 with the VLA by \citet{Puche1990} revealed only a small part of the outer disc, mainly due to the insufficient size of their mosaic of only five pointings.
  
\begin{figure}
    \centering
    \includegraphics[width=0.96\linewidth]{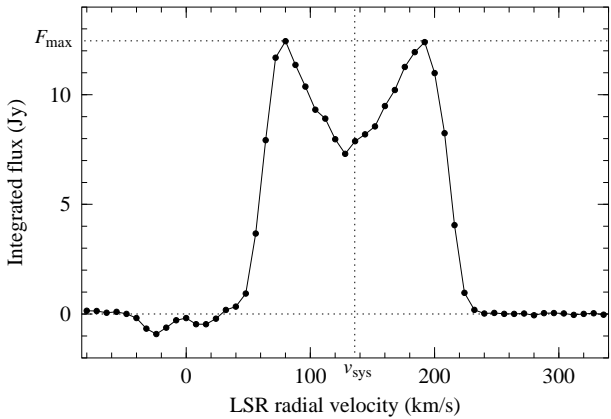}
    \caption{Integrated \ion{H}{i} spectrum of NGC~300. The fluctuations near $v_{\rm LSR} = 0~\mathrm{km \, s}^{-1}$ are due to Galactic foreground emission that was not deconvolved. The vertical dotted line indicates the systemic velocity of NGC~300 of $v_{\rm LSR} = 136~\mathrm{km \, s}^{-1}$ (equivalent to $144~\mathrm{km \, s}^{-1}$ in the barycentric reference frame).}
    \label{fig_intspec}
  \end{figure}

  The integrated \ion{H}{i} spectrum of NGC~300 is shown in Fig.~\ref{fig_intspec}. The total flux derived from our ATCA data is $1720~\mathrm{Jy \, km \, s}^{-1}$. The corresponding total \ion{H}{i} mass of NGC~300 amounts to $M_{\rm HI} = 1.5 \times 10^{9}~\mathrm{M}_{\odot}$ under the assumption that the gas is optically thin. Since this will not be the case in the dense inner regions of the disc, the derived mass must be considered a lower limit. In addition, some of the diffuse flux of NGC~300 will not have been picked up by the ATCA due to the lack of sufficiently short telescope baselines. The effect of missing short spacings and the inability of CLEAN to recover all the flux is visible in the channel maps of Fig.~\ref{fig_chanmaps}, with the \ion{H}{i} emission sitting in a slightly negative bowl.
  
  The \ion{H}{i} mass determined from the \ion{H}{i} Parkes All-Sky Survey (HIPASS; \citealt{Barnes2001}) is slightly larger with $M_{\rm HI} = (1.69 \pm 0.13) \times 10^{9}~\mathrm{M}_{\odot}$ \citep[scaled to a distance of $1.9~\mathrm{Mpc}$]{Koribalski2004}. This indicates that some flux is missing in our ATCA observations, although not much. On the other hand, HIPASS suffers from inaccurate bandpass calibration resulting in serious artefacts near bright and extended sources, and the true \ion{H}{i} mass of NGC~300 could even be somewhat higher than the one determined by \citet{Koribalski2004}. The total \ion{H}{i} mass of $8 \times 10^{8}~\mathrm{M}_{\odot}$ determined by \citet{Puche1990} from their VLA observations (again scaled to a distance of $1.9~\mathrm{Mpc}$) is significantly smaller, mainly reflecting the limited size of their mosaic on the sky and possibly the lack of short telescope baselines.
  
  We can also try to assess how the mass is distributed between the inner and outer disc of NGC~300. By simply defining a column density boundary of $N_{\rm HI} = 5 \times 10^{20}~\mathrm{cm}^{-2}$ between the two disc components we obtain \ion{H}{i} masses of $8.5 \times 10^{8}$ and $6.5 \times 10^{8}~\mathrm{M}_{\odot}$ for the inner and outer disc, respectively. Hence, a significant fraction of the \ion{H}{i} mass is within the extended outer disc.
  
\begin{figure}
    \centering
    \includegraphics[width=\linewidth]{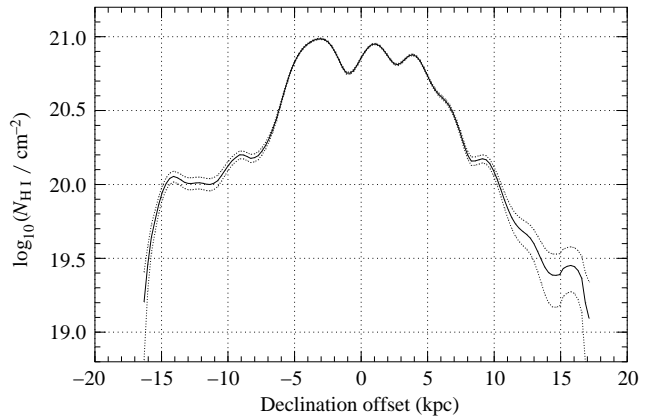}
    \caption{\ion{H}{i} column density profile of NGC~300 as a function of declination offset (solid curve). The uncertainties are plotted as the dotted curves and reflect our $5 \sigma$ \ion{H}{i} column density sensitivity of about $10^{19}~\mathrm{cm}^{-2}$.}
    \label{fig_nhiprofile}
  \end{figure}

  A closer inspection of the column density distribution reveals a certain degree of asymmetry in the disc. The south-eastern edge of the disc is rather clean and sharp, whereas the north-western edge looks more frayed and less well defined. This is illustrated in Fig.~\ref{fig_nhiprofile} showing the \ion{H}{i} column density profile across NGC~300 as a function of declination offset (i.e.\ along a line of constant right ascension passing through the centre of the galaxy). There is a distinctive plateau in the southern part of NGC~300 followed by a sudden and steep drop in column density towards the southern edge of the disc. In the northern part, in contrast, we observe a steady exponential decline without any obvious plateau or drop.
  
  The velocity dispersion map of NGC~300, derived from the fitting of Gauss--Hermite polynomials to the \ion{H}{i} spectra, is shown in the lower-right panel of Fig.~\ref{fig_coldens}. Velocity dispersions across the inner disc are variable and somewhat higher than in the outer disc with typical values in the range of about $10$ to $15~\mathrm{km \, s}^{-1}$. In contrast, the outer disc appears homogeneous with dispersions of just under $10~\mathrm{km \, s}^{-1}$ in most areas (equivalent to just over $20~\mathrm{km \, s}^{-1}$ FWHM for Gaussian line profiles). These values are not atypical for the warm neutral medium of the Milky Way \citep{Kalberla2009} and indicate an upper limit on the order of $10^{4}~\mathrm{K}$ for the gas temperature. The true gas temperature may be well below this upper limit, as effects such as turbulent motion and beam smearing will contribute to the observed line width.
  
\begin{figure*}
    \centering
    \includegraphics[width=0.8\linewidth]{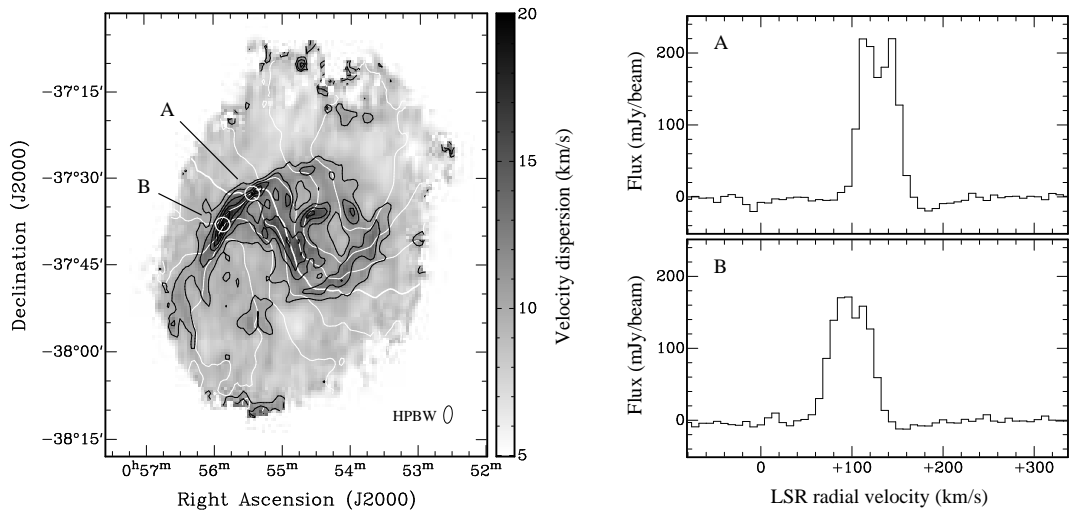}
    \caption{The greyscale image and black contour lines show the velocity dispersion of the \ion{H}{i} gas in NGC~300 as derived from Gauss--Hermite polynomials fitted to the spectral lines. The contour levels are $10$, $12$, $14$, $16$, $18$, and $20~\mathrm{km \, s}^{-1}$. The white contour lines show the velocity field for comparison and have been drawn in intervals of $20~\mathrm{km \, s}^{-1}$ centred about the systemic velocity of $v_{\rm LSR} = 136~\mathrm{km \, s}^{-1}$ (bold contour line; equivalent to $144~\mathrm{km \, s}^{-1}$ in the barycentric reference frame). The two spectra were extracted from the regions of highest dispersion marked with the white circles labelled A and B in the map.}
    \label{fig_dispersion}
  \end{figure*}

  Fig.~\ref{fig_dispersion} reveals regions of increased velocity dispersion near the boundary between the inner and outer \ion{H}{i} disc, in particular on the north-eastern side of the galaxy. The highest dispersions of about $20~\mathrm{km \, s}^{-1}$ are found along a filament in the eastern part of NGC~300. Such high values can no longer be explained by thermal line broadening alone, as in that case gas temperatures of about $5 \times 10^{4}~\mathrm{K}$ would be required. Instead, the high velocity dispersion suggests the presence of different gas components along the line of sight as supported by the spectra shown on the right-hand side of Fig.~\ref{fig_dispersion}. The spectra were extracted from the locations of highest dispersion marked with A and B in the map and reveal a clear double line profile at the boundary between the inner and outer \ion{H}{i} disc. Apparently, each disc component produces a separate line component in the spectrum, and the transition between the inner and outer disc is not gradual but abrupt within the size of our synthesized beam. This is also suggested by the radial velocity field as indicated by the white contours in Fig.~\ref{fig_dispersion}. In the transition region between the inner and outer disc we observe a strong discontinuity in the velocity field. The resulting kinks in the velocity contours coincide precisely with the regions of increased velocity dispersion in the map, thus confirming our impression of an abrupt transition between the inner and outer disc of NGC~300.
  
  Interestingly, the transition from the inner disc to the outer disc is less sharp on the south-western side of NGC~300 where the velocity contour lines are much smoother, and -- despite a slightly increased velocity dispersion -- the spectra do not exhibit a double-peak profile. Instead, there is a gradual transition from the inner to the outer disc component in the spectrum across several beam widths.
  
  \subsection{Rotation curve}
  \label{sect_rotcur}
  
  In order to determine the rotation curve, $v_{\rm rot}(r)$, of NGC~300 we must assume a model for the three-dimensional kinematics of the galaxy. One of the simplest and most commonly applied models is to assume that the gas particles move in circular orbits along tilted rings with varying position angle, $\varphi$, and inclination, $i$ \citep[e.g.][]{Rogstad1974}. In addition, each ring is assigned a position, $(x_{0},y_{0})$, and systemic velocity, $v_{\rm sys}$, both of which can also be allowed to vary.
  
  We used the GIPSY task \textsc{rotcur} to fit a tilted ring model to the radial velocity field of NGC~300 as derived from the fitting of Gauss--Hermite polynomials to the spectral lines. We fitted 20~rings with radii ranging from $r = 100$ to $2000~\mathrm{arcsec}$ ($0.9$ to $18.4~\mathrm{kpc}$) in steps of $100~\mathrm{arcsec}$. Each ring was chosen to be $100~\mathrm{arcsec}$ wide to match the angular resolution of our data. Data within an angle of $\pm 20^{\circ}$ around the minor axis of NGC~300 were excluded from the fit, and all data were weighted with the cosine of the position angle, $| \cos (\vartheta) |$.
  
  We first derived the position of the centre of NGC~300 from an optical $V$-band image taken with the MPG/ESO 2.2-m telescope at La Silla \citep{Pietrzynski2002}. NGC~300 has a very distinct, point-like core that is thought to be associated with either a nuclear stellar cluster or even a central black hole \citep{Soffner1996,Kim2004} and should therefore provide us with an accurate position of the dynamical centre. A Gaussian fit to the optical core revealed a central position of $\alpha = 00^{\rm h} 54^{\rm m} 53\fs{}4$ and $\delta = {-37}^{\circ} 41' 02\farcs{}6$ in the J2000.0 coordinate system.\footnote{This position is consistent with the dynamical centre of the innermost tilted rings as derived by \textsc{rotcur}.} Next, we ran \textsc{rotcur} with the optically derived central position fixed and the gas expansion velocity also fixed to a value of $v_{\rm exp} = 0~\mathrm{km \, s}^{-1}$. All other parameters (systemic velocity, rotation velocity, position angle, and inclination angle) were still left as free parameters. After the first run of \textsc{rotcur} we determined the systemic velocity by averaging over the fitted systemic velocities of all rings. The resulting value of $v_{\rm sys} = 136 \pm 2~\mathrm{km \, s}^{-1}$ in the LSR frame, corresponding to $144 \pm 2~\mathrm{km \, s}^{-1}$ in the barycentric reference frame, was then used and kept fixed in all subsequent runs. This result is consistent with the barycentric velocity of $v_{\rm sys} = 146 \pm 2~\mathrm{km \, s}^{-1}$ found by \citet{Koribalski2004} based on HIPASS.
  
  Next, we ran \textsc{rotcur} again with only the rotation velocity, position angle, and inclination angle left as free parameters. We followed the approach described by \citet{deBlok2008} and first fixed the radial position angle profile. To reduce the initial bumpiness of the position angle profile we smoothed it by applying a boxcar filter with a width of $300~\mathrm{arcsec}$ (equal to three radial bins) to the position angle solution. After the next run of \textsc{rotcur} we also smoothed and fixed the inclination angle profile using the same boxcar filter as for the position angle. With all other parameters fixed, we then ran \textsc{rotcur} for the last time to obtain the final solution for the rotation curve.
  
\begin{figure*}
    \centering
    \includegraphics[width=0.8\linewidth]{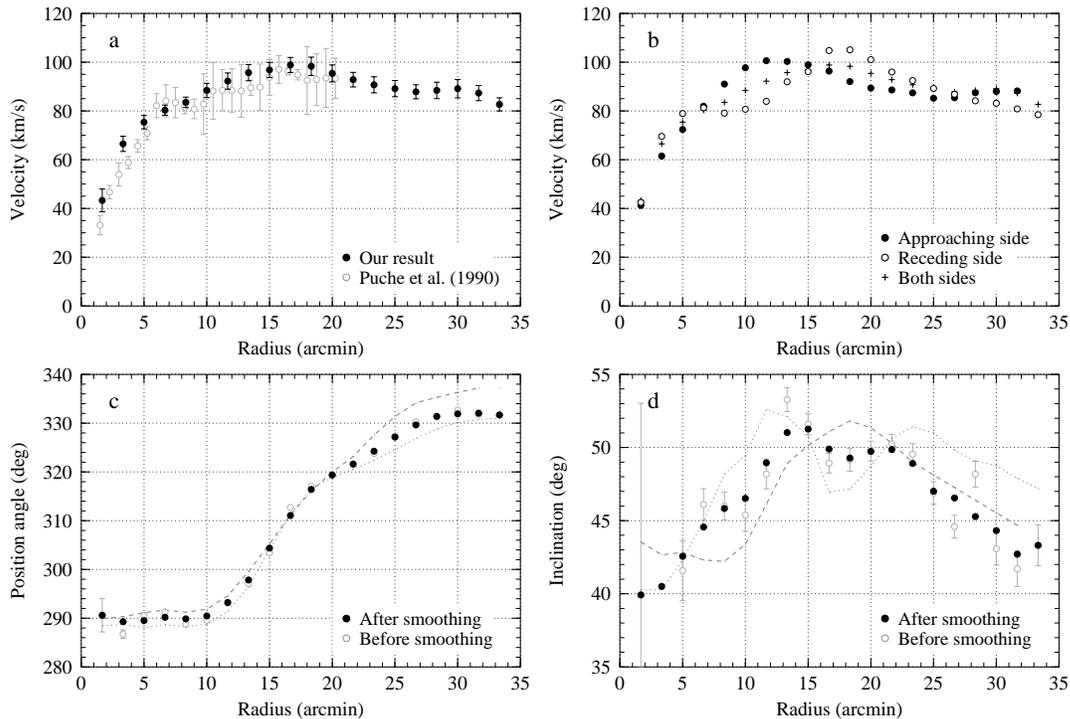}
    \caption{Panel~a shows the derived overall rotation curve of NGC~300 out to a radius of $2000~\mathrm{arcsec}$. The rotation curve determined by \citet{Puche1990} based on VLA data is shown for comparison. Panel~b shows the rotation curves derived from the approaching and receding side of NGC~300 separately compared to the overall rotation curve. Panels~c and~d show the position angle, $\varphi(r)$, and inclination angle, $i(r)$, respectively, of the tilted ring model used for the final rotation curve fit with \textsc{rotcur} (filled black circles) compared to the original values before smoothing (open grey circles with error bars). The dashed and dotted grey lines show the values derived separately for the approaching and receding side, respectively.}
    \label{fig_rotcur}
  \end{figure*}

\begin{figure*}
    \centering
    \includegraphics[width=\linewidth]{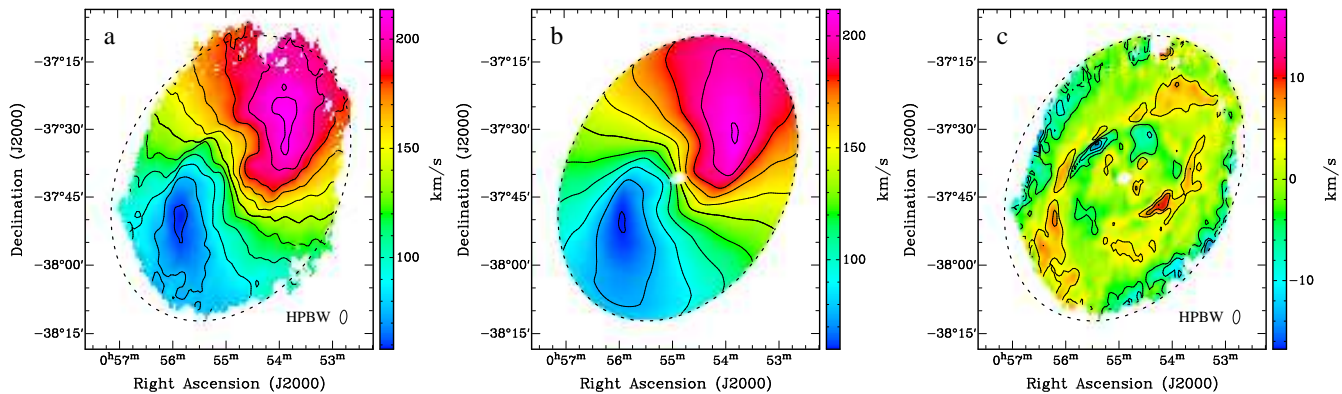}
    \caption{Panel~a shows the observed radial velocity field of NGC~300 based on the position of the intensity maximum of Gauss--Hermite polynomials fitted to the spectral line profiles. Panel~b shows the modelled radial velocity field derived from the overall rotation curve as shown in Fig.~\ref{fig_rotcur}. In both panels, the contour lines are drawn in intervals of $15~\mathrm{km \, s}^{-1}$ centred about the systemic velocity of $v_{\rm LSR} = 136~\mathrm{km \, s}^{-1}$ (bold contour line; equivalent to $144~\mathrm{km \, s}^{-1}$ in the barycentric reference frame). All velocities in the two maps are given in the LSR frame. Panel~c shows the difference between the observed and modelled velocity fields. The contour lines are drawn at levels of $\pm 4$, $\pm 8$, $\pm 12$, and $-16~\mathrm{km \, s}^{-1}$. The black, dotted ellipse in all three maps marks the outer edge of the model and illustrates the significant lopsidedness of the gas disc of NGC~300.}
    \label{fig_velofield}
  \end{figure*}

  Our final \ion{H}{i} rotation curve of NGC~300 is shown in Fig.~\ref{fig_rotcur}a and Table~\ref{tab_parameters}. A comparison between the observed and modelled velocity fields is presented in Fig.~\ref{fig_velofield}. Due to the combination of a large field of view with a fairly compact array configuration we are able to determine the rotation curve out to an angular radius of $2000~\mathrm{arcsec}$ equivalent to a physical radius of $18.4~\mathrm{kpc}$. This is a significant improvement compared to the previous measurement with the Very Large Array (VLA) by \citet{Puche1990} which covered a radius of only $11~\mathrm{kpc}$. The rotation curve derived by \citet{Puche1990} is also plotted in Fig.~\ref{fig_rotcur}a for comparison. Within their uncertainties both rotation curves agree quite well, although a direct comparison is difficult due to the different resolutions of the VLA observations and the ATCA observations presented here. At a higher resolution of $50~\mathrm{arcsec}$ FWHM, the rotation curve of \citet{Puche1990} shows small-scale variations not present in our ATCA rotation curve. In addition, there is some discrepancy in the very inner region of $r < 5~\mathrm{arcmin}$ which is most likely caused by different degrees of beam smearing in combination with a rapid change in rotation velocity in the inner part of the galaxy. Hence, our ATCA observations in particular are not capable of reliably defining the rotation curve of NGC~300 across the inner $3~\mathrm{kpc}$.
  
  The derived rotation curve of NGC~300 rises out to a radius of about $15$ to $20~\mathrm{arcmin}$ (equivalent to $8$ to $11~\mathrm{kpc}$) where we observe a maximum rotation velocity of $v_{\rm rot} = 98.8 \pm 3.3~\mathrm{km \, s}^{-1}$ at $r = 16\farcm{}7$ ($9.2~\mathrm{kpc}$). This value is consistent with the results of \citet{Puche1990} who found a maximum velocity of $v_{\rm rot} = 97.0 \pm 5.7~\mathrm{km \, s}^{-1}$ at a radius of $15\farcm{}8$ ($8.7~\mathrm{kpc}$). \citet{Rogstad1979} derived similar values in the range of $v_{\rm rot} = 94$ to $102~\mathrm{km \, s}^{-1}$ at a radius of $16~\mathrm{arcmin}$ ($8.8~\mathrm{kpc}$). Beyond about $10~\mathrm{kpc}$ the rotation velocity gradually decreases to below $85~\mathrm{km \, s}^{-1}$. This decrease was not seen in the previous VLA data of \citet{Puche1990} due to insufficient radial extent of their map.
  
  A particular problem of rotation curve fits in general is the determination of uncertainties. The statistical errors of the fit, as provided by \textsc{rotcur}, are usually much smaller than the actual uncertainties of the solution which are dominated by systematic errors such as non-circular motion of the observed gas, insufficient spatial resolution of the observations, effects of varying optical depth, etc. There are two possible ways to obtain more reliable estimates of the uncertainties. One possibility is to determine the standard deviation of velocities along each tilted ring from the mean value of the rotation velocity for that ring. This provides us with a more realistic assessment of the true deviations of the observed velocities from the model. The error bars of our final rotation curve presented in Fig.~\ref{fig_rotcur}a have been determined by this method.
  
  Another possibility is to derive the rotation curve for the approaching and receding side of the galaxy separately and assess the differences between the two curves. This is shown in Fig.~\ref{fig_rotcur}b where the approaching and receding curves are compared to the overall rotation curve. The fits for the approaching and receding sides are independent from the overall rotation curve fit, as we derived separate solutions for the position angle and inclination. The two rotation curves both show a clear maximum and then decrease again at larger radii. The position of the maximum as well as the overall shape of the rotation curve, however, differ substantially between the approaching and receding side. These discrepancies indicate systematic uncertainties of up to $\pm 10~\mathrm{km \, s}^{-1}$ for the overall rotation curve of NGC~300. The rotation curve of the receding side continually decreases after reaching a maximum at $r \approx 10~\mathrm{kpc}$, whereas the curve derived from the approaching side also decreases but then flattens out at larger radii. In addition, the approaching side has its rotation curve maximum significantly further inward at $r \approx 6.5~\mathrm{kpc}$. These differences indicate a high degree of asymmetry in the kinematics of the gas disc of NGC~300, some of which is immediately evident from the velocity map in the left-hand panel of Fig.~\ref{fig_velofield}. The results indicate that the finding of a decreasing rotation curve in the outer disc should be approached with caution because non-circular motions and asymmetries in the disc could have contributed significantly to the observed velocity field of NGC~300.
  
  Fig.~\ref{fig_rotcur}c and \ref{fig_rotcur}d show the position angle, $\varphi(r)$, and inclination, $i(r)$, used in the final rotation curve fit with \textsc{rotcur}. For comparison we also show the originally fitted values before applying the boxcar filter for smoothing and the separate solutions for the approaching and receding side of the galaxy. The position angle is constant over the entire inner disc of NGC~300 with $\varphi(r) = 290^{\circ}$ for $r \lesssim 10~\mathrm{arcmin}$. Beyond the edge of the inner disc the position angle of the gas disc gradually changes by about $40^{\circ}$ only to become constant again in the very outer disc with $\varphi(r) = 332^{\circ}$ for $r \gtrsim 30~\mathrm{arcmin}$. This behaviour of $\varphi(r)$ reflects the visual impression of a twist in the \ion{H}{i} disc of NGC~300 and suggests the presence of two distinct gas discs, each with its own spatial orientation. The variation in inclination, $i(r)$, with radius appears less regular. The inclination of the gas disc increases from about $40^{\circ}$ to just over $50^{\circ}$ within the inner $15~\mathrm{arcmin}$ ($8~\mathrm{kpc}$) and then gently decreases again to $i(r) \lesssim 45^{\circ}$ in the outermost regions of the disc.
  
  The observed decrease in rotation velocity is strongly coupled with the characteristics of the inclination angle in the outer parts of the disc. A smaller inclination angle would result in a higher rotation velocity. In order to obtain a flat rotation curve, the inclination angle of the outermost tilted ring would need to be about $35^{\circ}$. This is significantly smaller than the inclination angle of about $43^{\circ}$ resulting from the tilted ring model. At the same time, we estimate a geometric inclination angle of about $44^{\circ}$ by comparing the major and minor axis of the $2 \times 10^{19}~\mathrm{cm}^{-2}$ \ion{H}{i} column density contour level under the assumption of perfect circular symmetry of the disc. This is consistent with the kinematic inclination from the tilted ring model.
  
  It is interesting to note in this context that the maximum rotation velocity occurs near the transition radius between the inner and outer \ion{H}{i} disc. A closer inspection of the rotation curves derived for the approaching and receding sides of NGC~300 (Fig.~\ref{fig_rotcur}) reveals that the position of maximum rotation velocity is closely linked to strong variations in inclination angle at that particular radius. It is possible that part of this variation in inclination is an artefact caused by the failure of the tilted ring model to cope with the sudden transition between the inner and outer disc. Consequently, the decrease in rotation velocity across the outer disc of NGC~300 could be less pronounced than indicated by the tilted ring model.
  
  \section{Mass models}
  \label{sect_massmodels}
  
  After having determined the overall rotation curve, we can now try to derive mass models of NGC~300 using the GIPSY task \textsc{rotmas}. This will allow us to study the radial distribution of the different visible and dark matter components in NGC~300 and assess how much dark matter is required to explain the rotation curve. For this purpose, we assume the presence of three mass components, namely a gaseous disc, a stellar disc, and a dark matter halo. \textsc{rotmas} will attempt to fit the velocity curves of these three components to the total rotation curve:
  \begin{equation}
    v_{\rm rot}^{2}(r) = f_{\rm gas} v_{\rm gas}^{2}(r) + f_{\star} v_{\star}^{2}(r) + v_{\rm DM}^{2}(r) .
  \end{equation}
  The factors $f_{\star}$ and $f_{\rm gas}$ are mass scaling factors for the stellar and gaseous disc, respectively. \textsc{rotmas} determines the velocity curve, $v(r)$, of a mass component from its mass surface density, $\Sigma(r)$. In the following Sections~\ref{sect_gasmass} and~\ref{sect_starmass} we will derive the required mass surface densities of the gaseous and stellar disc. In Section~\ref{sect_darkmass} we will introduce the different dark matter models used in the mass modelling of NGC~300. Finally, Section~\ref{sect_massmodelling} will discuss the outcome of the mass modelling.
  
  \begin{table*}
    \caption{Modelling parameters of NGC~300. The columns denote: galactocentric radius, $r$, in arcsec and kpc; derived rotation velocity, $v_{\rm rot}$; tilted ring position angle, $\varphi$, and inclination, $i$; stellar mass surface density, $\Sigma_{\star}$, as derived from the \textit{Spitzer} IRAC $3.6~\mathrm{\umu{}m}$, $4.5~\mathrm{\umu{}m}$, and combined data; gas mass surface density, $\Sigma_{\rm gas}$.}
    \label{tab_parameters}
    \begin{tabular}{rrrrrrrrr}
      \hline
      $r$ & $r$ & $v_{\rm rot}$ & $\varphi$ & $i$ & $\Sigma_{\star}^{3.6 \, \mathrm{\umu{}m}}$ & $\Sigma_{\star}^{4.5 \, \mathrm{\umu{}m}}$ & $\Sigma_{\star}$ & $\Sigma_{\rm gas}$ \\
      ($\mathrm{arcsec}$) & (kpc) & ($\mathrm{km \, s}^{-1}$) & ($^{\circ}$) & ($^{\circ}$) & ($\mathrm{M}_{\odot} \, \mathrm{pc}^{-2}$) & ($\mathrm{M}_{\odot} \, \mathrm{pc}^{-2}$) & ($\mathrm{M}_{\odot} \, \mathrm{pc}^{-2}$) & ($\mathrm{M}_{\odot} \, \mathrm{pc}^{-2}$) \\
      \hline
      $ 100$ & $ 0.92$ & $43.3 \pm 4.7$ & $290.6$ & $39.9$ & $45.4 \pm 5.9$ & $41.8 \pm 6.1$ & $43.6 \pm 4.3$ & $6.3 \pm 0.8$ \\
      $ 200$ & $ 1.84$ & $66.5 \pm 3.1$ & $289.3$ & $40.5$ & $24.3 \pm 3.3$ & $22.8 \pm 3.5$ & $23.5 \pm 2.4$ & $7.4 \pm 1.2$ \\
      $ 300$ & $ 2.76$ & $75.4 \pm 2.7$ & $289.5$ & $42.6$ & $10.8 \pm 1.6$ & $10.1 \pm 1.7$ & $10.4 \pm 1.2$ & $7.0 \pm 1.6$ \\
      $ 400$ & $ 3.68$ & $80.3 \pm 2.1$ & $290.2$ & $44.6$ & $ 6.3 \pm 1.1$ & $ 6.2 \pm 1.2$ & $ 6.3 \pm 0.8$ & $7.1 \pm 1.7$ \\
      $ 500$ & $ 4.61$ & $83.5 \pm 2.2$ & $289.9$ & $45.8$ & $ 3.7 \pm 1.0$ & $ 3.7 \pm 0.9$ & $ 3.7 \pm 0.7$ & $7.2 \pm 1.5$ \\
      $ 600$ & $ 5.53$ & $88.4 \pm 2.9$ & $290.5$ & $46.5$ & $ 1.5 \pm 0.7$ & $ 1.7 \pm 0.6$ & $ 1.6 \pm 0.4$ & $6.8 \pm 2.0$ \\
      $ 700$ & $ 6.45$ & $92.2 \pm 3.3$ & $293.2$ & $49.0$ & $ 0.5 \pm 0.4$ & $ 0.5 \pm 0.5$ & $ 0.5 \pm 0.3$ & $4.8 \pm 1.8$ \\
      $ 800$ & $ 7.37$ & $95.7 \pm 3.3$ & $297.8$ & $51.0$ & $ 0.2 \pm 0.4$ & $ 0.1 \pm 0.6$ & $ 0.1 \pm 0.3$ & $3.5 \pm 1.5$ \\
      $ 900$ & $ 8.29$ & $96.8 \pm 3.1$ & $304.4$ & $51.3$ & $ 0.1 \pm 0.4$ & $-0.1 \pm 0.4$ & $ 0.0 \pm 0.3$ & $2.7 \pm 1.4$ \\
      $1000$ & $ 9.21$ & $98.8 \pm 3.1$ & $311.1$ & $49.9$ & $ 0.1 \pm 0.4$ & $-0.1 \pm 0.5$ & $ 0.0 \pm 0.3$ & $2.1 \pm 1.0$ \\
      $1100$ & $10.13$ & $98.3 \pm 3.8$ & $316.4$ & $49.3$ & $ 0.0 \pm 0.5$ & $-0.1 \pm 0.5$ & $ 0.0 \pm 0.3$ & $1.6 \pm 0.6$ \\
      $1200$ & $11.05$ & $95.4 \pm 3.5$ & $319.4$ & $49.7$ & $ 0.0 \pm 0.6$ & $ 0.1 \pm 0.5$ & $ 0.0 \pm 0.4$ & $1.5 \pm 0.6$ \\
      $1300$ & $11.98$ & $92.8 \pm 3.0$ & $321.6$ & $49.9$ & $-0.1 \pm 0.5$ & $ 0.0 \pm 0.4$ & $ 0.0 \pm 0.3$ & $1.3 \pm 0.5$ \\
      $1400$ & $12.90$ & $90.7 \pm 3.3$ & $324.2$ & $48.9$ & $ 0.0 \pm 0.5$ & $ 0.1 \pm 0.5$ & $ 0.0 \pm 0.4$ & $1.1 \pm 0.4$ \\
      $1500$ & $13.82$ & $89.1 \pm 3.3$ & $327.1$ & $47.0$ &        --      &        --      &        --      & $0.8 \pm 0.3$ \\
      $1600$ & $14.74$ & $87.8 \pm 2.9$ & $329.6$ & $46.6$ &        --      &        --      &        --      & $0.6 \pm 0.3$ \\
      $1700$ & $15.66$ & $88.4 \pm 3.4$ & $331.4$ & $45.3$ &        --      &        --      &        --      & $0.4 \pm 0.2$ \\
      $1800$ & $16.58$ & $89.1 \pm 3.8$ & $331.9$ & $44.3$ &        --      &        --      &        --      & $0.3 \pm 0.1$ \\
      $1900$ & $17.50$ & $87.3 \pm 3.1$ & $332.0$ & $42.7$ &        --      &        --      &        --      & $0.2 \pm 0.1$ \\
      $2000$ & $18.42$ & $82.7 \pm 2.7$ & $331.7$ & $43.3$ &        --      &        --      &        --      & $0.1 \pm 0.1$ \\
      \hline
    \end{tabular}
  \end{table*}
  
  \subsection{Gaseous component}
  \label{sect_gasmass}
  
  Deriving the radial mass distribution of the gaseous disc in NGC~300 is relatively straightforward. We first used the GIPSY task \textsc{ellint} to derive the radial \ion{H}{i} column density profile from the column density map, using the tilted ring parameters obtained from the rotation curve fit with \textsc{rotcur}. The column density profile, $N_{\rm H\,I}(r)$, can then be converted into gas mass surface density via
  \begin{equation}
    \Sigma_{\rm gas}(r) = f \mathrm{m}_{\rm H} N_{\rm H\,I}(r) \cos(i)
  \end{equation}
  where $\mathrm{m}_{\rm H} = 1.674 \times 10^{-27}~\mathrm{kg}$ is the mass of a hydrogen atom, and $f$ is a mass correction factor. For our analysis we assume a value of $f = 1.4$ to account for primordial helium expected to be associated with the neutral hydrogen gas in the disc of NGC~300. The factor $\cos(i)$ is required to de-project all column densities to face-on values. The resulting mass surface density profile of the gas disc is shown in Fig.~\ref{fig_massprofile} and Table~\ref{tab_parameters}.
  
  Throughout the inner part of the gas disc the surface density is basically constant with $\Sigma_{\rm gas} \approx 7~\mathrm{M}_{\odot} \, \mathrm{pc}^{-2}$ (including the correction for helium). Beyond a radius of about $6~\mathrm{kpc}$ the gas mass surface density begins to decline. The outer profile can be reasonably well described by either an exponential decline, $\Sigma_{\rm gas} \propto \exp(-r / h)$, with a radial scale length of $h = 3.44 \pm 0.15~\mathrm{kpc}$, or by a power law, $\Sigma_{\rm gas} \propto (r / \mathrm{kpc})^{a}$, with a slope of $a = -2.38 \pm 0.07$.
  
  This method, of course, assumes that the \ion{H}{i} gas is optically thin which might not be the case for the densest regions along the spiral arms of NGC~300. As a consequence, column densities, and hence gas mass surface densities, in these regions would be underestimated. In addition, we have made the assumption that the gas disc is infinitely thin, and any significant thickness of the disc would result in errors in the determination of \ion{H}{i} column densities, depending also on the inclination angle of the disc.
  
  \subsection{Stellar component}
  \label{sect_starmass}
  
  Deriving the stellar mass distribution in NGC~300 is significantly more complicated. In order to determine the radial mass profile of the stellar component of NGC~300 and its contribution to the derived rotation curve we need to find a way to convert the optical or near-infrared flux density profile, $S_{\lambda}(r)$, into stellar mass surface density, $\Sigma_{\star}(r)$, via the stellar mass-to-light ratio, $\Upsilon_{\lambda}$, such that
  \begin{equation}
    \Sigma_{\star}(r) \sim \Upsilon_{\lambda} \, S_{\lambda}(r) . \label{eqn_surface}
  \end{equation}
  The index $\lambda$ indicates that both the mass-to-light ratio as well as the flux density are wavelength-dependent and are usually determined within a specific photometric band of wavelength $\lambda$.
  
  By using near-infrared instead of optical data one can largely overcome the problem of absorption by dust in the measurement of the radial flux density profile of the stellar component. NGC~300 has been observed with the Infrared Array Camera (IRAC) on board the \textit{Spitzer Space Telescope} \citep{Helou2004}, and for this work we have used IRAC data in the $3.6$ and $4.5~\mathrm{\umu m}$ bands to determine the stellar surface brightness profile. The nominal wavelengths of these two bands are $3.550$ and $4.493~\mathrm{\umu m}$, respectively \citep{Fazio2004}. IRAC pipeline images are usually calibrated in terms of flux density in units of $\mathrm{MJy \, sr}^{-1}$.
  
  The flux density profile, $S_{\lambda}(r)$, in the $3.6$ and $4.5~\mathrm{\umu m}$ wavebands must then be converted to stellar mass density, $\Sigma_{\star}(r)$, using the stellar mass-to-light ratio in units of the solar value, $\Upsilon_{\lambda}' \equiv \Upsilon_{\lambda} / \Upsilon_{\lambda}^{\odot}$. The corresponding conversion formula in Eq.~\ref{eqn_surface} can be rewritten in the form
  \begin{equation}
    \Sigma_{\star}(r) = C_{\lambda} \Upsilon_{\lambda}' \frac{f_{\rm A} S_{\lambda}(r)}{S_{0}} \label{eqn_sigma}
  \end{equation}
  where $C_{\lambda}$ is a wavelength-dependent conversion factor, $f_{\rm A}$ is the wavelength-dependent aperture correction factor for extended sources in IRAC images, and $S_{0}$ denotes the wavelength-dependent zero-magnitude flux. Based on the calculations of \citet{Oh2008} we derive conversion factors in the $3.6$ and $4.5~\mathrm{\umu m}$ wavebands of
  \begin{gather}
    C_{3.6} = 0.196~\mathrm{M}_{\odot} \, \mathrm{pc}^{-2} , \\
    C_{4.5} = 0.201~\mathrm{M}_{\odot} \, \mathrm{pc}^{-2} .
  \end{gather}
  According to \citet{Reach2005}, the aperture correction factors for the $3.6$ and $4.5~\mathrm{\umu m}$ bands are $f_{\rm A} = 0.944$ and $0.937$, respectively, and the corresponding zero-magnitude fluxes are $S_{0} = 280.9 \pm 4.1$ and $179.7 \pm 2.6~\mathrm{Jy}$.
  
  Before we can use Eq.~\ref{eqn_sigma} to convert radial flux density profiles into stellar mass density profiles we first need to determine the stellar mass-to-light ratio in NGC~300. Determination of the stellar mass-to-light ratio is very difficult because it depends on several uncertain parameters such as the initial mass function (IMF) of NGC~300, its star formation history, metallicity, etc. To make matters worse, most of these parameters will change with growing distance from the centre of a galaxy, resulting in a radial variation, $\Upsilon_{\lambda}(r)$, of the stellar mass-to-light ratio.
  
  Colour indices provide a convenient way to determine the mass-to-light ratio through basic photometric measurements. \citet{Bell2001} used galaxy evolution models to explore the relation between stellar mass-to-light ratio and optical/near-infrared colours of galaxies. They found a strong and robust correlation between the stellar mass-to-light ratio and the integrated colour of the stellar population in their model galaxies. They also concluded that a modified \citet{Salpeter1955} IMF with a smaller fraction of low-mass stars is consistent with the observed stellar mass-to-light ratios of a sample of spiral galaxies used for comparison with the models. From the relations between mass-to-light ratio and colour found by \citet{Bell2001} we derive the following relation between the $K$-band stellar mass-to-light ratio, $\Upsilon_{K}'$, and the $J - K$ colour index:
  \begin{equation}
    \log_{10} \left( \Upsilon_{K}' \right) = 1.434 \, (J - K) - 1.380 . \label{eqn_upsilonk}
  \end{equation}
  The relation was derived using the mass-dependent formation epoch model with bursts and the modified Salpeter IMF adopted by \citet{Bell2001}. Since our surface brightness measurements are based on \textit{Spitzer} data, we have to convert $\Upsilon_{K}'$ into the stellar mass-to-light ratios derived for the \textit{Spitzer} IRAC bands of $3.6$ and $4.5~\mathrm{\umu m}$. Using stellar population synthesis models, \citet{Oh2008} found the following well-defined correlations:
  \begin{gather}
    \Upsilon_{3.6}' = 0.92 \, \Upsilon_{K}' - 0.05 , \label{eqn_upsilon36} \\
    \Upsilon_{4.5}' = 0.91 \, \Upsilon_{K}' - 0.08 . \label{eqn_upsilon45}
  \end{gather}
  With the set of equations derived above we now have all tools at hand to determine the stellar mass surface density of NGC~300.
  
  \begin{figure}
    \centering
    \includegraphics[width=\linewidth]{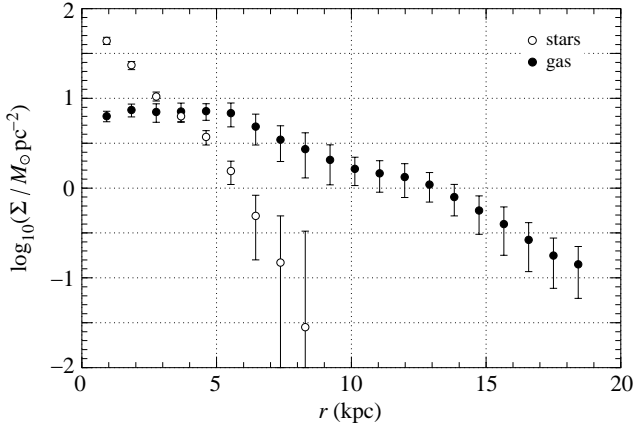}
    \caption{Azimuthally averaged radial profiles of the stellar mass surface density, $\Sigma_{\star}(r)$, and the gas mass surface density, $\Sigma_{\rm gas}(r)$, of NGC~300 (see Table~\ref{tab_parameters}).}
    \label{fig_massprofile}
  \end{figure}
  
  From the Two Micron All Sky Survey (2MASS) Large Galaxy Atlas \citep{Jarrett2003} we derive a global colour index of $J - K = 0.646 \pm 0.033~\mathrm{mag}$ for NGC~300. Inserting this value into Eq.~\ref{eqn_upsilonk} yields a stellar mass-to-light ratio for NGC~300 in the $K$-band of
  \begin{equation}
    \log_{10} \left( \Upsilon_{K}' \right) = {-0.454} \pm 0.047 .
  \end{equation}
  Using Eq.~\ref{eqn_upsilon36} and~\ref{eqn_upsilon45}, we derive stellar mass-to-light ratios in the IRAC $3.6$ and $4.5~\mathrm{\umu m}$ bands of
  \begin{gather}
    \log_{10} \left( \Upsilon_{3.6}' \right) = {-0.564} \pm 0.055 , \\
    \log_{10} \left( \Upsilon_{4.5}' \right) = {-0.620} \pm 0.054 .
  \end{gather}
  We can now use Eq.~\ref{eqn_sigma} to derive the stellar mass surface density of NGC~300. Again, we used the GIPSY task \textsc{ellint} with the tilted ring fitting parameters derived earlier to determine the surface brightness profiles of NGC~300 from the \textit{Spitzer} $3.6$ and $4.5~\mathrm{\umu m}$ images. Both profiles were then corrected for inclination and combined into a single stellar mass surface density profile which is shown in Fig.~\ref{fig_massprofile} and Table~\ref{tab_parameters}. The resulting total stellar mass is $(1.0 \pm 0.1) \times 10^{9}~\mathrm{M}_{\odot}$.
  
  The derived profile can be described by an exponential decline, $\Sigma_{\star}(r) \propto \exp(-r/h)$, with a radial scale length of $h = 1.39 \pm 0.04~\mathrm{kpc}$. This value is in good agreement with the scale length of $1.38~\mathrm{kpc}$ measured in the $I$ band by \citet{Kim2004}. Comparison with the values of $2.06~\mathrm{kpc}$ in the $B_{J}$ band \citep{Carignan1985} and $1.94~\mathrm{kpc}$ in the $B$ band \citep{Kim2004} suggests that the outer stellar disc of NGC~300 is bluer than the inner disc.
  
  \begin{table*}
    \centering
    \caption{Results of the mass modelling of NGC~300 for the pseudo-isothermal (ISO), Burkert, and NFW dark matter halo profiles. $f_{\star}$ and $f_{\rm gas}$ are the mass scaling factors for the stellar and gaseous disc, respectively. The parameters in columns~4 and~5 are the free model parameters of the different dark matter models as explained in the text. Column~6 lists the reduced $\chi^{2}$ value of the model as a measure of the goodness of fit. The last two columns list the total mass, $M_{\rm tot}$, of the model within a radius of $r = 18.4~\mathrm{kpc}$ and the fraction of dark matter, $f_{\rm DM}$. Note that the NFW fit with fixed gas mass produced negative stellar masses and is therefore not listed.}
    \label{tab_massmodels}
    \begin{tabular}{lrrrrrrr}
      \hline
      Halo model          &   $f_{\star}$ & $f_{\rm gas}$ &     $r_{\rm c}$ & $\varrho_{0}$ & $\chi_{\rm red}^{2}$ & $M_{\rm tot}$ & $f_{\rm DM}$ \\
                          &               &               &           (kpc) & ($\mathrm{M}_{\odot} \, \mathrm{pc}^{-3}$) &  & ($10^{10} \, \mathrm{M}_{\odot}$) &              \\
      \hline
      ISO (fixed)         &           $1$ &           $1$ & $0.93 \pm 0.14$ & $0.170 \pm 0.046$ &           $2.10$ & $3.6$ & $0.92$ \\
      ISO (gas fixed)     & $1.6 \pm 1.3$ &           $1$ & $1.17 \pm 0.61$ & $0.107 \pm 0.111$ &           $2.16$ & $3.6$ & $0.90$ \\
      ISO (free)          & $2.3 \pm 0.6$ & $4.8 \pm 1.0$ & $0.74 \pm 0.35$ & $0.135 \pm 0.119$ &           $1.17$ & $3.3$ & $0.65$ \\
      \hline
                          &   $f_{\star}$ & $f_{\rm gas}$ &     $r_{\rm c}$ & $\varrho_{0}$ & $\chi_{\rm red}^{2}$ & $M_{\rm tot}$ & $f_{\rm DM}$ \\
                          &               &               &           (kpc) & ($\mathrm{M}_{\odot} \, \mathrm{pc}^{-3}$) &  & ($10^{10} \, \mathrm{M}_{\odot}$) & \\
      \hline
      Burkert (fixed)     &           $1$ &           $1$ & $2.78 \pm 0.14$ & $0.081 \pm 0.008$ &           $1.18$ & $3.2$ & $0.91$ \\
      Burkert (gas fixed) & $2.2 \pm 0.9$ &           $1$ & $3.58 \pm 0.75$ & $0.044 \pm 0.021$ &           $1.20$ & $3.3$ & $0.88$ \\
      Burkert (free)      & $3.1 \pm 0.9$ & $3.0 \pm 1.4$ & $4.04 \pm 1.10$ & $0.025 \pm 0.018$ &           $1.12$ & $3.3$ & $0.73$ \\
      \hline
                          &   $f_{\star}$ & $f_{\rm gas}$ &     $r_{\rm s}$ &     $r_{200}$ & $\chi_{\rm red}^{2}$ & $M_{\rm tot}$ & $f_{\rm DM}$ \\
                          &               &               &           (kpc) &             (kpc) &                  & ($10^{10} \, \mathrm{M}_{\odot}$) & \\
      \hline
      NFW (fixed)         &           $1$ &           $1$ & $5.81 \pm 0.56$ &    $89.7 \pm 2.4$ &           $1.47$ & $3.4$ & $0.91$ \\
      NFW (free)          & $1.1 \pm 1.1$ & $3.9 \pm 1.0$ & $4.00 \pm 1.53$ &    $71.9 \pm 6.7$ &           $0.95$ & $3.3$ & $0.74$ \\
      \hline
    \end{tabular}
  \end{table*}
  
  As mentioned earlier, we cannot expect the stellar mass-to-light ratio to be constant across the entire galaxy, but there will likely be a radial variation in $\Upsilon_{\lambda}$ reflecting different star formation histories and stellar populations at different distances from the centre of NGC~300. A way to account for this radial variability of $\Upsilon_{\lambda}$ would be to determine the radial profile of the $J - K$ colour index of NGC~300 based on infrared imaging data. Unfortunately, the signal-to-noise ratio in the 2MASS images is very low, and determination of $\Upsilon_{\lambda}$ as a function of galactocentric distance is therefore not feasible due to large uncertainties.
  
  \subsection{Dark matter component}
  \label{sect_darkmass}
  
  \subsubsection{Pseudo-isothermal halo}
  
  The spherical pseudo-isothermal halo is one of the simplest models for the density profile of dark matter haloes. Its density profile resembles that of an isothermal sphere and reads
  \begin{equation}
    \varrho(r) = \frac{\varrho_{0}}{1 + (r / r_{\rm c})^{2}}
  \end{equation}
  \citep[e.g.][]{Begeman1991} where $\varrho_{0}$ is the central density of the halo and $r_{\rm c}$ the so-called core radius. The corresponding velocity profile is
  \begin{equation}
    v^{2}(r) = 4 \upi \mathrm{G} \varrho_{0} r_{\rm c}^{2} \left[ 1 - \frac{r_{\rm c}}{r} \arctan \left( \frac{r}{r_{\rm c}} \right) \right]
  \end{equation}
\citep{Kent1986}.
  
  \subsubsection{Burkert halo}
  
  For his study of dwarf galaxy rotation curves \citet{Burkert1995} introduced the following empirical halo density profile:
  \begin{equation}
    \varrho(r) = \frac{\varrho_{0} r_{\rm c}^{3}}{(r + r_{\rm c}) (r^{2} + r_{\rm c}^{2})} .
  \end{equation}
  Again, $\varrho_{0}$ and $r_{\rm c}$ are the central density and core radius of the halo, respectively. The resulting velocity profile reads
  \begin{equation}
    \begin{split}
      v^{2}(r) & = \frac{6.4 \mathrm{G} \varrho_{0} r_{\rm c}^{3}}{r} \left[ \ln \left( 1 + \frac{r}{r_{\rm c}} \right) \right. \\
      & \qquad \left. + \frac{1}{2} \ln \left( 1 + \frac{r^{2}}{r_{\rm c}^{2}} \right) - \arctan \left( \frac{r}{r_{\rm c}} \right) \right]
    \end{split}
  \end{equation}
  \citep{Salucci2000}. The major difference between the Burkert halo and the pseudo-isothermal halo is that for large radii the Burkert halo density behaves as $\varrho(r) \sim r^{-3}$, whereas for the pseudo-isothermal halo we get a different behaviour of $\varrho(r) \sim r^{-2}$.
  
  \subsubsection{NFW halo}
  \label{sect_nfw}
  
  The NFW dark matter density profile was suggested by \citet{Navarro1995,Navarro1996} based on numerical simulations of dark matter haloes in a hierarchically clustering universe. They found that in their models the radial density profile of dark matter haloes can be accurately described by
  \begin{equation}
    \varrho(r) = \frac{\delta_{\rm c} \varrho_{\rm crit}}{\frac{r}{r_{\rm s}} \left( 1 + \frac{r}{r_{\rm s}} \right)^{2}}
  \end{equation}
  where $\varrho_{\rm crit}$ is the critical density of the universe, $\delta_{\rm c}$ is the so-called characteristic, dimensionless density of the halo, and $r_{\rm s}$ is a scale radius. According to \citet{Navarro1997}, the resulting circular velocity curve of the dark matter halo reads
  \begin{equation}
    v^{2}(r) = \frac{v_{200}^{2}}{x} \, \frac{\ln(1 + cx) - \frac{cx}{1 + cx}}{\ln(1 + c) - \frac{c}{1 + c}}
  \end{equation}
  with the substitution $x = r / r_{200}$ and the concentration parameter $c = r_{200} / r_{\rm s}$ (not to be confused with the speed of light, $\mathrm{c}$). In this form of the equation, $r_{200}$ denotes the radius within which the mean density of the halo equals 200 times the critical density of the universe, $\varrho_{\rm crit} = 3 \mathrm{H}_{0}^{2} / (8 \upi \mathrm{G})$. This choice is somewhat arbitrary, although $r_{200}$ corresponds to approximately the virial radius of the halo. Accordingly, $v_{200}$ is the circular velocity of the halo at $r_{200}$ expressed by
  \begin{equation}
    v_{200} = \sqrt{\frac{\mathrm{G} M_{200}}{r_{200}}} = 10 \mathrm{H}_{0} r_{200} .
  \end{equation}
  This leaves us with only two free parameters for the rotation curve fit, namely $r_{\rm s}$ and $r_{200}$. In addition, we assumed a Hubble constant of $\mathrm{H}_{0} = 70.5~\mathrm{km} \, \mathrm{s}^{-1} \, \mathrm{Mpc}^{-1}$ \citep{Hinshaw2009} throughout this paper. Note that the NFW density profile has a singularity at the centre of the halo. Nevertheless, both the mass and circular velocity are well-defined and finite at all radii.
  
  \subsection{Mass modelling of NGC~300}
  \label{sect_massmodelling}
  
  Throughout the literature there has been no consistent approach as to what vertical density profile, $\varrho(z)$, to assume for the gaseous and stellar discs. \citet{Capozziello2007}, for example, assumed both discs to be infinitely thin, whereas \citet{deBlok2008} assumed an infinitely thin gas disc but a $\mathrm{sech}^{2}(z/z_{0})$ distribution for the stellar disc. \citet{Sanders2007}, in turn, also assumed the gas disc to be extended in the vertical direction with the same thickness as that of the stellar disc. In contrast to all these approaches, \citet{Kalberla2007} demonstrated that the \ion{H}{i} disc of the Milky Way shows strong flaring, with the scale height (half width at half maximum) increasing from $z_{0} = 60~\mathrm{pc}$ at a radius of $4~\mathrm{kpc}$ up to $z_{0} \approx 2.7~\mathrm{kpc}$ at a radius of $40~\mathrm{kpc}$.
  
  The $\mathrm{sech}^{2}(z/z_{0})$ distribution of the stellar disc is motivated by studies of edge-on spiral galaxies by \citet{vanderKruit1981a,vanderKruit1981b}. They found that the mass density of the stellar disc is of the form
  \begin{equation}
    \varrho(r,z) = \varrho(r) \, \mathrm{sech}^2 (z / z_{0})
  \end{equation}
  with the scale height, $z_{0}$, being largely independent of radius, $r$ (but see \citet{deGrijs1997} for evidence of a variation of $z_{0}$ with $r$). We adopt this distribution in our model of NGC~300 and assume a ratio of $h / z_{0} = 5$ in accordance with previous studies \citep[e.g.][]{vanderKruit1981a}. From the exponential scale length, $h = 1.39 \pm 0.04~\mathrm{kpc}$, of the combined $3.6$ and $4.5~\mathrm{\umu m}$ mass surface density profile we therefore derive a vertical scale height of $z_{0} \approx 280~\mathrm{pc}$.
  
  As for the gaseous disc, we have considered two cases, namely an infinitely thin disc and an exponential distribution with a scale height equivalent to that of the stellar disc. It turns out that a thicker gas disc results in a slightly lower rotation velocity and a general smoothing of the velocity profile of the gaseous component. However, the introduction of a non-zero scale height for the gaseous disc does not have any significant influence on the mass modelling as long as the scale height is not larger than a few hundred $\mathrm{pc}$. We therefore decided to make the assumption of an infinitely thin gas disc in all our models.
  
  \begin{table}
    \centering
    \caption{Results of the mass modelling of NGC~300 for Newtonian dynamics and Modified Newtonian Dynamics (MOND) under the assumption of no additional dark matter halo. $f_{\star}$ and $f_{\rm gas}$ are the mass scaling factors for the stellar and gaseous disc, respectively. The last column lists the reduced $\chi^{2}$ value of the model as a measure for the goodness of fit.}
    \label{tab_massmodels2}
    \begin{tabular}{lrrr}
      \hline
      Dynamics              &     $f_{\star}$ &   $f_{\rm gas}$ & $\chi_{\rm red}^{2}$ \\
      \hline
      Newtonian (gas fixed) & $9.5 \pm 1.4$ &           $1$ & $67.42$ \\
      Newtonian (free)      & $4.2 \pm 0.3$ & $8.4 \pm 0.3$ &  $2.17$ \\
      \hline
      MOND (gas fixed)      & $2.2 \pm 0.2$ &           $1$ &  $3.31$ \\
      MOND (free)           & $2.6 \pm 0.2$ & $0.6 \pm 0.2$ &  $2.50$ \\
      \hline
    \end{tabular}
  \end{table}
  
\begin{figure*}
    \centering
    \includegraphics[width=0.95\linewidth]{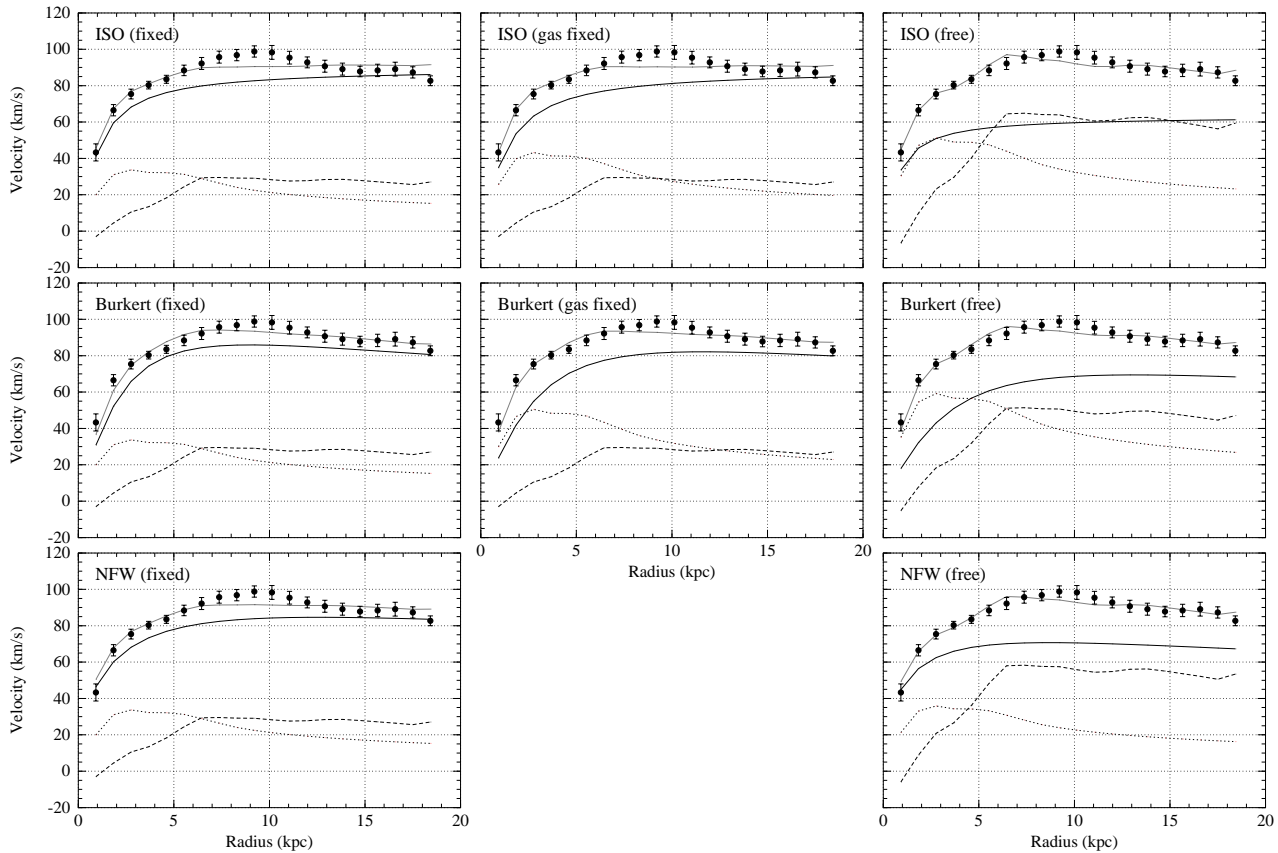}
    \caption{Results of the fitting of mass models to the rotation curve of NGC~300. The panels show, from top to bottom, the results for the pseudo-isothermal halo, the Burkert halo, and the NFW halo. For each halo model the results for fixed, partly fixed (gas only), and free stellar and gaseous mass components is shown. The black data points with error bars show the observed rotation curve. The stellar, gaseous, and dark matter components are plotted as the dotted, dashed, and solid black lines, respectively. The total velocity profile of the model is plotted as the solid grey line. Note that the NFW fit with fixed gas mass produced negative stellar masses and is therefore not displayed.}
    \label{fig_massmodels}
  \end{figure*}

  For all three halo models we carried out fits for fixed, partly fixed (gas only), and variable stellar and gaseous mass components.\footnote{Note that the fit for the NFW halo with fixed gas mass scaling factor failed, because the best-fitting model resulted in a negative stellar mass scaling factor. Therefore, this case will not be presented and discussed here.} The resulting mass model fits are shown in Table~\ref{tab_massmodels} and Fig.~\ref{fig_massmodels}. All three dark matter halo models result in reasonably good fits. The goodness of fit of the pseudo-isothermal halo model is somewhat lower compared to the Burkert and NFW halo models. The lower performance of the pseudo-isothermal halo is mainly due to its $r^{-2}$ density profile at large radii. As a consequence, the velocity curve will continue to increase and eventually converge to a parameter-dependent constant, $v_{\infty}$. Therefore, the pseudo-isothermal halo model cannot explain the decreasing rotation curve of NGC~300, whereas the Burkert and NFW halo models (both with an $r^{-3}$ behaviour) are doing better. Of course, all three halo models produce a lower reduced $\chi^{2}$ in the case of free stellar and gas mass scaling factors. Whereas for the pseudo-isothermal and NFW haloes the improvement is significant, the Burkert halo is the only model to produce a good fit with fixed mass scaling factors. In the case of free mass scaling factors we obtain values in the range of $f_{\star} \approx 1 \ldots 3$ and $f_{\rm gas} \approx 3 \ldots 5$. Fixing the gas mass scaling factor to $f_{\rm gas} = 1$ yields somewhat smaller stellar mass scaling factors of $f_{\star} \approx 1.6 \ldots 2.2$.
  
  As the stellar mass scaling factor strongly depends on the assumptions made about the stellar mass-to-light ratio of NGC~300, a higher value of $f_{\star}$ can arise from differences in star formation history or metallicity compared to the standard spiral galaxy model found by \citet{Bell2001}. In comparison, the best mass model fit for NGC~300 found by \citet{Kent1987} yields $f_{\star} = 1.0$ (relative to our stellar disc mass of $1.0 \times 10^{9}~\mathrm{M}_{\odot}$), and his maximum-disc model has $f_{\star} = 3.1$. \citet{Puche1990} derive somewhat larger values of $f_{\star} = 2.2$ and $4.3$ for their best-fitting and maximum-disc models, respectively. These numbers are all in general agreement, but they also expose the uncertainties in the determination of the stellar mass-to-light ratio and the total stellar disc mass of NGC~300.
  
  The free mass scaling factors for the gaseous component are somewhat higher than expected. A factor of~2 can still be explained by optical depth effects in the inner regions of the gas disc, significant amounts of ionised gas throughout the disc, or missing flux in our \ion{H}{i} map due to lack of short interferometer spacings. Gas mass scaling factors of~4 or~5, however, seem unreasonably large and not realistic.
  
  Note that in the very inner region of NGC~300 the value of the rotation velocity of the gas disc is negative (see Fig.~\ref{fig_massmodels}). This does not imply counter-rotation of the gas or the presence of negative mass, but simply reflects a net outward force on test particles due to the central depression of the gas disc. This net outward force results in a negative value of $v_{\rm gas}^{2}$ and therefore an imaginary velocity in the mathematical description of mass models that is usually expressed by negative values of $v_{\rm gas}$.
  
\begin{figure*}
    \centering
    \includegraphics[width=0.7\linewidth]{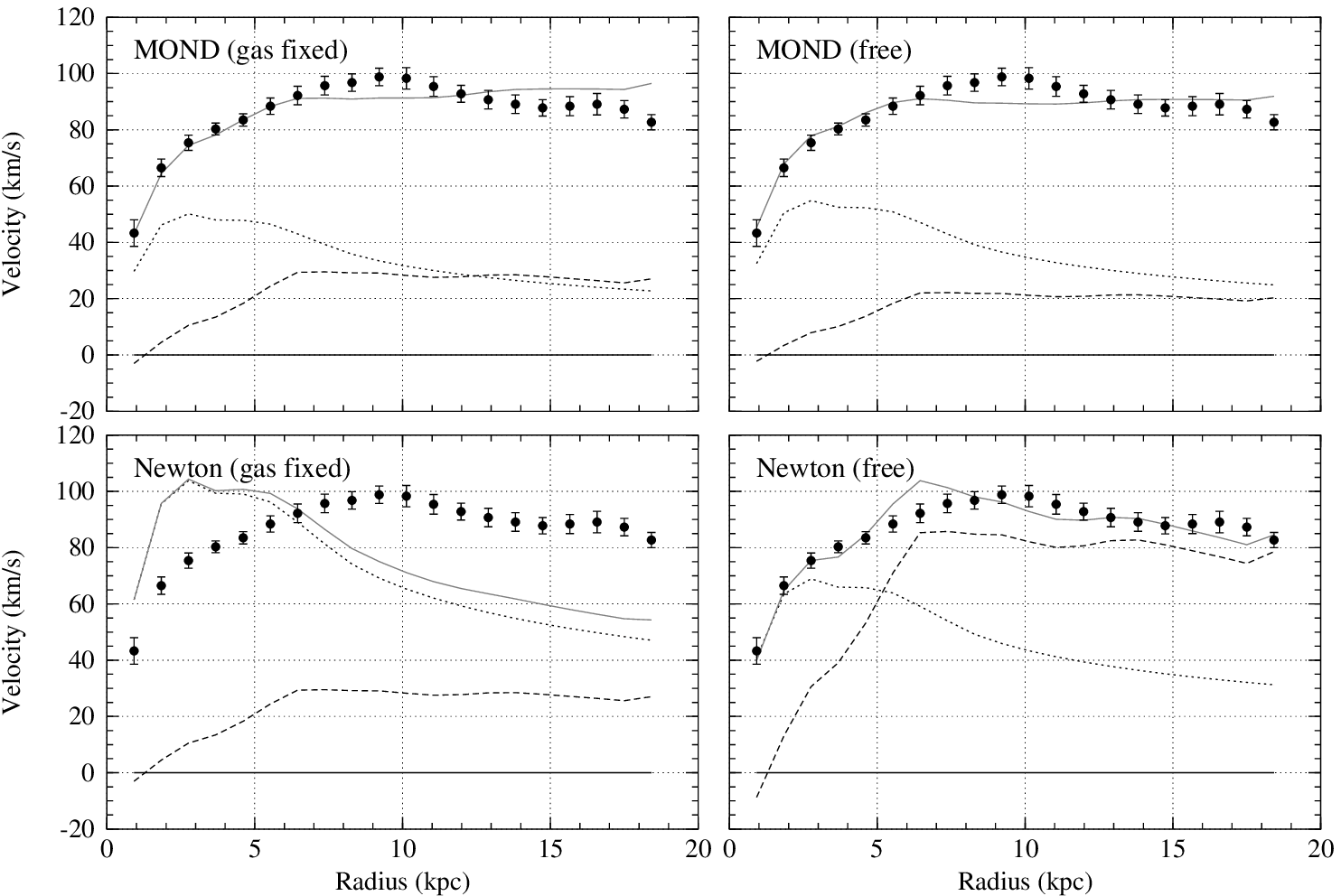}
    \caption{Same as in Fig.~\ref{fig_massmodels}, but for non-relativistic MOND (top) and Newtonian dynamics (bottom). In both cases, no additional dark matter was included. In the left-hand column the gaseous mass component was held fixed whereas the stellar component was allowed to vary. In the right-hand column both components were variable.}
    \label{fig_massmodels2}
  \end{figure*}

  In addition to the dark matter models discussed above we also fitted two models without any dark matter component at all, using Newtonian dynamics and (non-relativistic) Modified Newtonian Dynamics (MOND; \citealt{Milgrom1983}), respectively. In both cases, only the stellar and gas mass scaling factors were left as free parameters of the fitting procedure. For the MOND model we assumed a constant $a_{0} = 1.21 \times 10^{-10}~\mathrm{m \, s}^{-2}$ \citep{Begeman1991}.\footnote{Fits with variable $a_{0}$ and fixed mass scaling factors resulted in very similar values of $a_{0} \simeq 1.3 \times 10^{-10}~\mathrm{m \, s}^{-2}$.} The resulting mass models are shown in Table~\ref{tab_massmodels2} and Fig.~\ref{fig_massmodels2}.
  
  Neither of the two models produces meaningful results. The best fit in the MOND scenario ($\chi_{\rm red}^{2} = 2.50$) yields mass scaling factors of $f_{\star} = 2.6 \pm 0.2$ and $f_{\rm gas} = 0.6 \pm 0.2$ for the stellar and gaseous disc, respectively. Apart from the significantly larger $\chi_{\rm red}^{2}$ compared to all of the dark matter models with free mass scaling factors, the MOND model results in a considerable downscaling of the gas disc mass. This is an implausible result because the derived gas mass surface density was directly derived from the observed \ion{H}{i} column densities and must therefore be a lower limit. Fixing the mass scaling factor for the gas disc to a value of $f_{\rm gas} = 1$ slightly reduces the stellar mass scaling factor to $f_{\star} = 2.2 \pm 0.2$, the goodness-of-fit parameter, however, significantly deteriorates to $\chi_{\rm red}^{2} = 3.31$.
  
  The reason for the inferior performance of MOND is the decrease in rotation velocity at larger radii. MOND has been designed to explain the flat rotation curves originally observed in many galaxies. Therefore, the model will fail in cases of decreasing rotation curve. Consequently, deep \ion{H}{i} observations with the aim to detect the faint, outer gas discs of galaxies have the potential to challenge MOND through the detection of decreasing rotation curves in the outer regions of many more galaxies. It should be noted, though, that -- as discussed earlier -- the observed decrease in the rotation velocity of NGC~300 is uncertain due to the possibility of non-circular motions and asymmetries in the disc. Deep studies of a much larger sample of spiral galaxies will be required to investigate the frequency of occurrence of decreasing rotation curves in these galaxies.
  
  The Newtonian scenario without dark matter also fails to produce meaningful results. The best fit again has a large $\chi_{\rm red}^{2} = 2.17$. In addition, both the stellar and gaseous disc masses would have to be scaled up significantly by scaling factors of $f_{\star} = 4.2 \pm 0.3$ and $f_{\rm gas} = 8.4 \pm 0.3$, respectively. Such large scaling factors, in particular for the gas disc, are highly implausible. Of course, fixing the gas mass scaling factor to $f_{\rm gas} = 1$ does not produce any meaningful result at all, because the remaining free stellar mass component dramatically fails to reproduce the overall shape of the rotation curve, resulting in a $\chi_{\rm red}^{2}$ of $67.4$.
  
  In summary, we obtain the best mass model fits under the assumption of a Burkert or NFW dark matter halo. Both dark matter models can cope with the decreasing rotation curve of NGC~300 at larger radii. The Burkert halo provides the best fit with fixed mass scaling factors, but both halo models produce similarly good fits with variable mass scaling factors, although the gas mass scaling factor is slightly larger and less realistic for the NFW halo. The resulting total mass of NGC~300 within $18.4~\mathrm{kpc}$ amounts to $(2.9 \pm 0.2) \times 10^{10}~\mathrm{M}_{\odot}$. For fixed mass scaling factors almost 90~per cent of the mass is contributed by dark matter. In the case of variable mass scaling factors we obtain 68~per cent dark matter for both the Burkert and NFW halo models (but only 57~per cent for the pseudo-isothermal halo model).
  
  However, all these results should be approached with great caution. First of all, none of the dark matter models discussed above is physically motivated. Instead, these are heuristic models based mainly on the results of numerical computer simulations. Secondly, the lower $\chi_{\rm red}^{2}$ values of the models with dark matter compared to those without dark matter may simply reflect the larger number of free fitting parameters in the former case. Any increase in the number of mass components or free parameters may naturally improve the goodness of fit, no matter what dark matter model one assumes.
  
  \section{Discussion}
  \label{sect_discussion}
  
  \subsection{Origin of the outer disc}
  \label{sect_outerdisc}
  
  A conspicuous feature of NGC~300 is its extended outer \ion{H}{i} disc spanning more than $1^{\circ}$ (equivalent to about $35~\mathrm{kpc}$) across the sky. Although the outer disc was partly mapped by previous observations (e.g., \citealt{Puche1990}), this is the first time that its extent has been imaged out to the $10^{19}~\mathrm{cm}^{-2}$ column density level with moderately high spatial resolution of approximately $1~\mathrm{kpc}$. The results of our tilted ring model suggest that there is a substantial change in position angle between the inner and outer disc, resulting in a twisted appearance of NGC~300 which is particularly obvious in the velocity field.
  
  One possible scenario for this twist is that the distortion and warping of the outer disc of NGC~300 was caused by tidal forces during a recent encounter with another galaxy. A potential candidate for a close encounter could be NGC~55. Both galaxies are separated by about $8^{\circ}$ on the sky, corresponding to a projected separation of about $270~\mathrm{kpc}$.\footnote{NGC~55 has approximately the same distance from the Milky Way as NGC~300 \citep{Pietrzynski2006}.} Under the assumption that both galaxies move away from each other at a relative velocity of $200~\mathrm{km \, s}^{-1}$, there would have been a close encounter between NGC~55 and NGC~300 about $1.3~\mathrm{Ga}$ ago. This time scale is comparable to the $1.5~\mathrm{Ga}$ that have passed since the previous perigalactic passage of the Magellanic Clouds on their orbit about the Milky Way and the resulting creation of the Magellanic Stream as predicted by numerical simulations (\citealt{Gardiner1996,Yoshizawa2003}; however see \citealt{Mastropietro2009}).
  
  Alternatively, the distortion of the \ion{H}{i} disc of NGC~300 could have been caused by a recent encounter with a smaller companion galaxy. The nearest currently known companion is the dwarf-spheroidal galaxy ESO~294$-$G010. Its distance of about $1.9~\mathrm{Mpc}$ \citep{Karachentsev2003} is the same as that of NGC~55 and NGC~300, and its angular separation from NGC~300 is $6\fdg{}9$, equivalent to a physical separation of about $230~\mathrm{kpc}$. Its small size \citep{Bouchard2005} and relatively large separation from NGC~300, however, suggest that ESO~294$-$G010 has never had any significant gravitational influence on NGC~300. At the same time, ESO~294$-$G010 is much closer to NGC~55 with an angular separation of only $3\fdg{}5$ equivalent to about $115~\mathrm{kpc}$.
  
  Deep optical observations of two fields in the outer part of NGC~300 with the Gemini South 8-m telescope by \citet{Bland-Hawthorn2005} revealed an extended stellar disc reaching out to a radius of at least $24~\mathrm{arcmin}$ (equivalent to about $14~\mathrm{kpc}$ or 10~optical scale lengths). Surprisingly, the luminosity of the stellar disc of NGC~300 shows a simple exponential profile over its entire radial extent (except for the nuclear region) without any indication of a deviation or break. This seems to contradict our observation of a strongly warped \ion{H}{i} disc with a rather discontinuous transition between the inner and outer disc. If this warp were the result of tidal interaction, the stellar disc should have been affected in the same way.
  
  It is important to note that \citet{Bland-Hawthorn2005} did not map the entire two-dimensional distribution of the outer disc on the sky, but observed only two fields in the south-eastern quadrant of NGC~300. Any warp in the stellar disc would therefore have to affect the stars at this particular azimuthal angle to be discernible, and further optical imaging at different azimuthal angles would be required to firmly exclude the presence of a warp in the outer stellar disc.
  
  \citet{Bland-Hawthorn2005} discuss some of the implications of this discrepancy and speculate about the origin of the extended stellar disc, including the possibility of stars being scattered from the inner disc. Another possibility is that gas accreted by NGC~300 through hot or cold accretion modes could have fuelled low levels of star formation in the outer disc. In this case, however, there should be a break in the radial stellar surface brightness profile.
  
  Alternatively, stars could have formed in the outer \ion{H}{i} disc as a result of instabilities such as spiral density waves. In order to test this hypothesis, \citet{Bland-Hawthorn2005} calculated the Toomre $Q$ parameter,
  \begin{equation}
    Q(r) = \frac{\sigma(r) \kappa(r)}{3.36 \, \mathrm{G} \, \Sigma_{\rm gas}(r)} \label{eqn_toomreq}
  \end{equation}
  \citep{Toomre1964}, based on the \ion{H}{i} data of \citet{Puche1990}. Here, $\sigma(r)$ is the velocity dispersion of the gas disc, $\kappa(r)$ is the epicyclic frequency of the rotating disc defined by
  \begin{equation}
    \kappa^{2} = 2 \left( \frac{v^2}{r^2} + \frac{v}{r} \frac{\mathrm{d}v}{\mathrm{d}r} \right) ,
  \end{equation}
  and $\Sigma_{\rm gas}(r)$ denotes the gas mass surface density, all three of which are a function of galactocentric radius, $r$. \citet{Bland-Hawthorn2005} found values in excess of $Q \approx 5$ for the disc beyond a radius of $10~\mathrm{arcmin}$ (equivalent to about $5.5~\mathrm{kpc}$), suggesting that the outer disc of NGC~300 is stable with respect to axisymmetric instabilities.
  
\begin{figure}
    \centering
    \includegraphics[width=\linewidth]{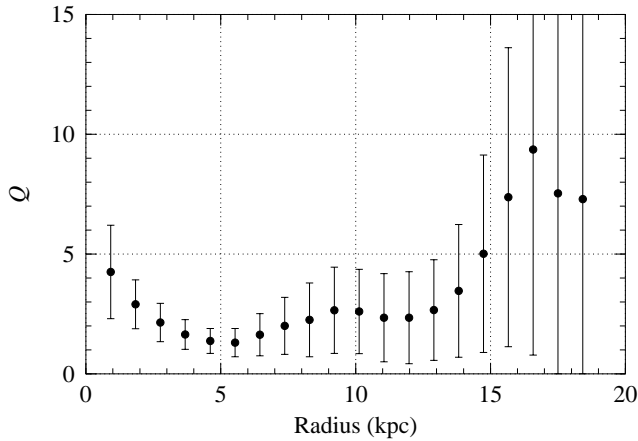}
    \caption{Toomre $Q$ parameter of the gas disc of NGC~300 as a function of galactocentric radius.}
    \label{fig_toomreq}
  \end{figure}

  Recalculation of $Q$ based on our \ion{H}{i} observations with the ATCA, however, suggests somewhat smaller $Q$ parameters over much of the disc of NGC~300. Using Eq.~\ref{eqn_toomreq} and assuming the same constant gas velocity dispersion of $\sigma = 5~\mathrm{km \, s}^{-1}$ as \citet{Bland-Hawthorn2005} yields typical values of $Q \approx 1 \ldots 3$ between $2$ and $12~\mathrm{kpc}$ radius (Fig.~\ref{fig_toomreq}), suggesting that the gas disc may be susceptible to axisymmetric instabilities. However, a slightly higher and more realistic velocity dispersion of $10~\mathrm{km \, s}^{-1}$ \citep{Tamburro2009} would result in the disc being Toomre stable at most radii. Furthermore, recent studies suggest that the star formation efficiency in disc galaxies is regulated by other effects, such as the physics of the ISM on small scales \citep{Leroy2008} or the gas pressure in the disc \citep{Blitz2006}, and that the Toomre $Q$ parameter does not necessarily provide an accurate description of the ability of the disc to form stars.
  
  \begin{figure*}
    \centering
    \includegraphics[width=0.8\linewidth]{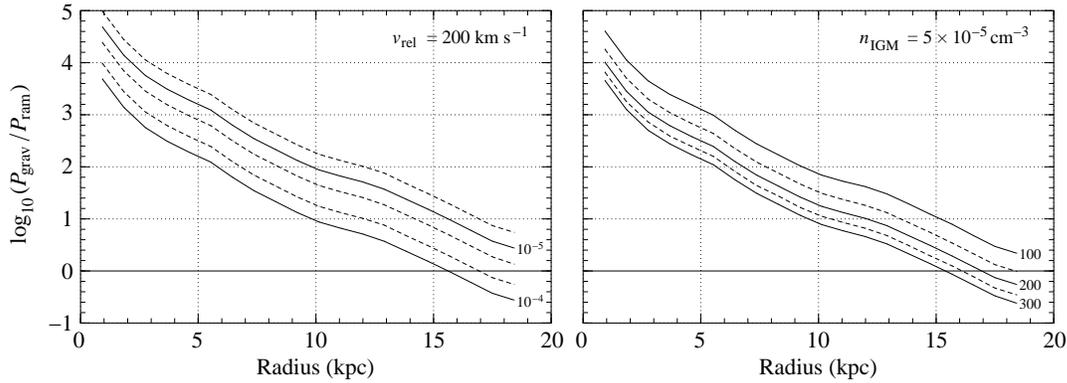}
    \caption{Ratio between gravitational pressure and ram pressure in the disc of NGC~300 as a function of radius. The left-hand panel shows the pressure ratio for a constant velocity of $v_{\rm rel} = 200~\mathrm{km \, s}^{-1}$ and five different densities of (from top to bottom) $n_{\rm IGM} = 5 \times 10^{-6}$, $1 \times 10^{-5}$, $2 \times 10^{-5}$, $5 \times 10^{-5}$, and $1 \times 10^{-4}~\mathrm{cm}^{-3}$. The right-hand panel shows the pressure ratio for a constant density of $n_{\rm IGM} = 5 \times 10^{-5}~\mathrm{cm}^{-3}$ and five different velocities of (from top to bottom) $v_{\rm rel} = 100$, $150$, $200$, $250$, and $300~\mathrm{km \, s}^{-1}$. The region above the horizontal black line at $\log_{10}(P_{\rm grav} / P_{\rm ram}) = 0$ is dominated by gravity, and the gas disc is stable. The region below that line is dominated by ram pressure, and the gas will be subject to ram pressure stripping.}
    \label{fig_rampressure}
  \end{figure*}
  
  Obviously, the next logical step in our attempt to understand the evolutionary history of NGC~300 would be a more detailed study of the outer stellar disc to determine the age and chemical composition of the stars. This was recently achieved by \citet{Vlajic2009} through $g'$ and $i'$ photometry of three fields in the outskirts of NGC~300 with the Gemini South telescope. The observations confirmed that the stellar population in the outer disc is predominantly old. At the same time, there appears to be a radial gradient in metallicity with a sudden break at a radius of about $10~\mathrm{kpc}$. While the inner disc shows a radially decreasing metallicity, this trend surprisingly reverses in the outer disc. Interestingly, the break occurs near the boundary between the inner and outer \ion{H}{i} disc of NGC~300, although it is not clear at this point whether this is just a coincidence or whether there is a physical connection between the two phenomena.
  
  \citet{Vlajic2009} propose two possible scenarios for the observed break in the metallicity gradient. Either there could have been some kind of radial mixing of stars in the disc, or star formation could have slowly progressed outward during the evolution of NGC~300, resulting in a moderate metallicity gradient in the inner disc, but an almost flat and pristine metallicity distribution across the outer disc. In both cases, the gas in the outer disc would not play any significant role in the recent star formation activity of NGC~300, which is consistent with the generally old age ($> 1~\mathrm{Ga}$) of the stellar population in the outer disc and the presence of only a small fraction of younger (a few hundred million years) main sequence stars.
  
  \subsection{Ram pressure interaction}
  \label{sect_rampressure}
  
  The aforementioned strong asymmetries found in the outer \ion{H}{i} disc of NGC~300 are suggestive of ram pressure interaction with the surrounding intergalactic medium (IGM) of the Sculptor Group. As NGC~300 is moving through the IGM, gas at the leading edge of the disc would be compressed, whereas gas at the trailing edge of the disc would be dragged and stripped by ram pressure forces (see, e.g., \citealt{Vollmer2009} and references therein). The influence of ram pressure has been mainly observed in the dense intergalactic medium of massive galaxy clusters \citep[e.g.][]{White1991,Vollmer2004}, in particular the nearby Virgo cluster \citep{Giovanelli1983}, where ram pressure is believed to be responsible for the stripping and removal of gas from galaxies \citep{Vollmer2001}. In contrast, only very few cases of ram pressure stripping in galaxy groups have been reported so far \citep[e.g.,][]{Kantharia2005,McConnachie2007,Bouchard2007}.
  
  To assess whether ram pressure could have caused the observed asymmetries in the gas disc of NGC~300 we can compare the expected ram pressure forces with the gravitational forces stabilising the disc. The pressure imposed on the gas disc by a galaxy's own gravitational potential can be expressed as
  \begin{equation}
    P_{\rm grav} = \Sigma_{\rm gas}(r) \left| \frac{\partial \Phi(r)}{\partial z} \right|_{\rm max} \label{eqn_gravity}
  \end{equation}
  \citep{Gunn1972,Roediger2005}. Here, $\Sigma_{\rm gas}(r)$ is the gas mass surface density of the disc as a function of radius, and $\Phi(r)$ denotes the gravitational potential of the galaxy. The term on the right-hand side of Eq.~\ref{eqn_gravity} determines the maximum of the absolute value of the gravitational acceleration perpendicular to the disc, with $z$ being the height above the disc.
  
  The gas disc will be subject to ram pressure stripping if the pressure exerted by the IGM exceeds the stabilising gravitational pressure as determined by Eq.~\ref{eqn_gravity}. The ram pressure of the IGM is determined by the density of the IGM, $\varrho_{\rm IGM}$, and the relative velocity of the galaxy with respect to the surrounding medium, $v_{\rm rel}$, thus
  \begin{equation}
    P_{\rm ram} = \varrho_{\rm IGM} v_{\rm rel}^{2} . \label{eqn_rampressure}
  \end{equation}
  Ram-pressure stripping of gas in the disc will occur whenever the instability condition $P_{\rm ram} > P_{\rm grav}$ is fulfilled. As the gravitational pressure will be a strong function of galactocentric radius, one can also define a truncation radius at which ram pressure forces start to dominate over gravity. The simple analytic approach described above assumes that the galaxy is moving face-on through the IGM, however, hydrodynamical simulations have demonstrated that even for larger inclination angles of the disc of up to about $60^{\circ}$ with respect to the direction of motion the analytic approach produces meaningful results \citep{Roediger2006}.
  
  In order to assess the stability of the gas disc of NGC~300 against ram pressure stripping we have calculated the instability condition for different scenarios. We assumed the gravitational potential of the galaxy to be dominated by its dark matter halo and neglected the potential of the stellar disc and gas disc. This assumption will be met at greater heights, $z$, above the disc plane in the outermost regions of the disc where we expect the gas to be particularly vulnerable to ram pressure stripping. Furthermore, we assumed the dark matter halo to have an NFW profile with the parameters (for the model with fixed stellar and gaseous mass scaling factors) derived in Section~\ref{sect_massmodelling} and listed in Table~\ref{tab_massmodels}. The gravitational potential of the NFW halo reads
  \begin{equation}
    \Phi_{\rm NFW}(r) = -\frac{\mathrm{G} M_{200} \ln \! \left( 1 + \frac{r}{r_{\rm s}} \right)}{r \left[ \ln(1 + c) - \frac{c}{(1 + c)} \right]}
  \end{equation}
  \citep{Hayashi2007} where $M_{200}$, $r_{\rm s}$, and $c$ are parameters of the NFW profile as discussed in Section~\ref{sect_nfw}. Solving Eq.~\ref{eqn_gravity} for the NFW potential and combining it with Eq.~\ref{eqn_rampressure} allows us to investigate the instability condition for different values of the IGM density and relative velocity of NGC~300. The results are presented in Fig.~\ref{fig_rampressure}.
  
  \begin{figure*}
    \centering
    \includegraphics[width=0.85\linewidth]{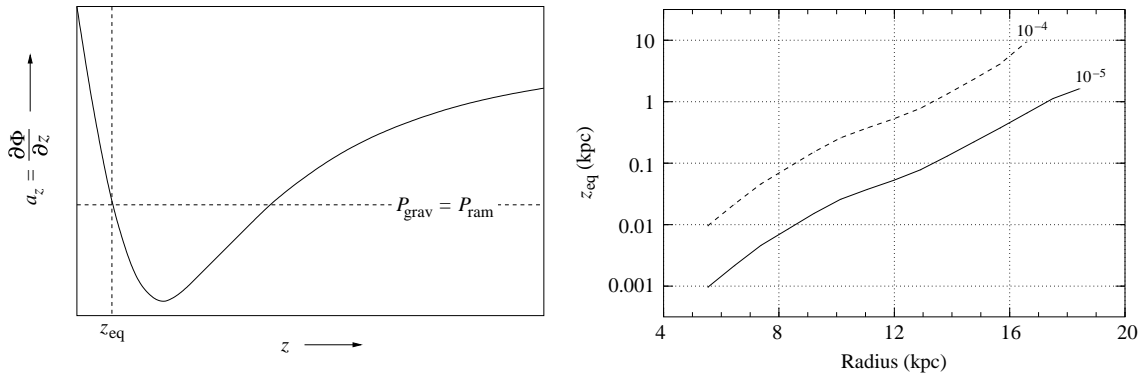}
    \caption{Left: Schematic overview of the gravitational acceleration, $a_{z}$, of the NFW halo perpendicular to the disc as a function of height, $z$, above the plane. At $z = z_{\rm eq}$, gravitational pressure and ram pressure are in balance. Right: Height, $z_{\rm eq}$, above the disc plane of NGC~300 where equilibrium between gravitational pressure and ram pressure is achieved. The situation for two different IGM densities of $n_{\rm IGM} = 10^{-5}$ and $10^{-4}~\mathrm{cm}^{-3}$ is shown under the assumption of face-on motion and a relative velocity of NGC~300 with respect to the surrounding IGM of $v_{\rm rel} = 150~\mathrm{km \, s}^{-1}$.}
    \label{fig_rampressure2}
  \end{figure*}
  
  The left-hand panel of Fig.~\ref{fig_rampressure} shows the situation for a constant velocity of $v_{\rm rel} = 200~\mathrm{km \, s}^{-1}$ of NGC~300 with respect to the IGM but for five different IGM densities. The right-hand panel of Fig.~\ref{fig_rampressure} shows the situation for a constant IGM density of $n_{\rm IGM} = 5 \times 10^{-5}~\mathrm{cm}^{-3}$ but for five different velocities. As we would expect, most of the gas disc of NGC~300 is many orders of magnitude above the instability condition of $\log_{10}(P_{\rm grav} / P_{\rm ram}) < 0$ represented by the horizontal black line. However, under certain conditions the outermost part of the gas disc beyond a radius of about $15~\mathrm{kpc}$ will get near or just below the instability line, in particular for IGM densities in excess of about a few times $10^{-5}~\mathrm{cm}^{-3}$ and sufficiently high relative velocities of $v_{\rm rel} \gtrsim 200~\mathrm{km \, s}^{-1}$. The density values assumed here are consistent with constraints on the density of the Local Group medium by various studies \citep[e.g.,][]{Rasmussen2001,Sembach2003,Williams2005}.
  
  It is interesting to note that the outer edge of the \ion{H}{i} disc of NGC~300 appears to be close to balance between gravitational and ram pressure forces over a range of reasonable values for density and velocity. This could indicate that the edge of the disc is defined by the aforementioned truncation radius at which ram pressure forces start to dominate over gravity. A stable disc configuration beyond that radius would be impossible, suggesting that under certain conditions ram pressure could in general be responsible for establishing the outer edge of \ion{H}{i} discs of galaxies in group environments.
  
  Furthermore, ram pressure could also result in a significant vertical displacement of disc gas and thereby contribute to the warping observed in the outer disc of many spiral galaxies. This situation is shown schematically in the left-hand panel of Fig.~\ref{fig_rampressure2} where the gravitational acceleration, $a_{z}$, of the NFW halo perpendicular to the gas disc is shown as a function of height, $z$, above the disc plane. Under the assumption of a face-on wind caused by the surrounding IGM, there will be for each galactocentric radius an equilibrium height, $z_{\rm eq}$, at which the gravitational pressure in the disc will be in balance with the ram pressure of the wind. Of course, there will be no such equilibrium for radii greater than the truncation radius beyond which the ram pressure always exceeds the gravitational pressure at any height above the disc.
  
  The right-hand panel of Fig.~\ref{fig_rampressure2} shows the expected value of $z_{\rm eq}$ as a function of galactocentric radius in NGC~300 for two different IGM densities of $n_{\rm IGM} = 10^{-5}$ and $10^{-4}~\mathrm{cm}^{-3}$ and under the assumption of a constant relative velocity of $v_{\rm rel} = 150~\mathrm{km \, s}^{-1}$. Over much of the disc of NGC~300 the expected equilibrium height, and thus vertical displacement, of the gas is very small. Towards the outer disc the equilibrium height exceeds $1~\mathrm{kpc}$ with a maximum value of $z_{\rm eq} \approx 1.6~\mathrm{kpc}$ in the outermost radial bin under the assumption of $n_{\rm IGM} = 10^{-5}~\mathrm{cm}^{-3}$. If we increase the IGM density by a factor of ten to $10^{-4}~\mathrm{cm}^{-3}$ the equilibrium height increases to almost $10~\mathrm{kpc}$ at $r = 16.6~\mathrm{kpc}$, and beyond that radius there is no solution at all because we would exceed the truncation radius. These estimates illustrate that ram pressure may have an impact on the vertical structure of the \ion{H}{i} of NGC~300 depending on the properties of the surrounding IGM. We can also rule out higher IGM densities of well above $10^{-5}~\mathrm{cm}^{-3}$ as they would result in a destruction of the outer parts of the observed \ion{H}{i} disc.
  
  Please note that in all our calculations we neglected the gravitational potential of the stellar and gaseous discs, hence underestimating the restoring force. This will particularly be the case for the inner disc of NGC~300 and locations close to the disc plane, where the ratio between gravitational pressure and ram pressure will be significantly higher than indicated in Fig.~\ref{fig_rampressure}. Near the outer edge of the \ion{H}{i} disc, however, the contribution from stars and gas to the gravitational potential is much smaller compared to the dark matter component. While the stellar mass surface density across the outer disc is immeasurably small, the contribution of the gas disc to the gravitational potential is comparable to that of the dark matter halo within a few hundred $\mathrm{pc}$ of the disc plane and negligible for $z \gtrsim 1~\mathrm{kpc}$. Therefore, our general conclusions about the stability of the outer disc are not affected by the additional contribution from the gas, and our numerical results near the outer edge of the disc are accurate within a factor of about $2$ to $3$.
  
  Our results suggest that the edge of the gas disc of NGC~300 may indeed be subject to ram pressure stripping, although the effect will likely be subtle. This result is consistent with our \ion{H}{i} observations which show notable asymmetries between the north-western and south-eastern part of NGC~300 and signs of distortion along the north-western edge of the \ion{H}{i} disc. In this scenario the proper motion of NGC~300 on the sky would be in the south-eastern direction, pointing away from NGC~55.
  
  The ram-pressure scenario is also consistent with the offsets between the \ion{H}{i} gas and the stellar distribution found in several dwarf galaxies of the Sculptor group by \citet{Bouchard2005}.\footnote{It should be noted, though, that their ATCA detections are of low significance, resulting in large uncertainties in the position of the \ion{H}{i} emission.} Although these effects are by far not as strong as in massive galaxy clusters, our results suggest that even in modest-sized galaxy groups ram pressure interaction does play a role in shaping the faint, outer gas discs of spiral galaxies in a density regime often not exposed by shallow \ion{H}{i} observations. This finding highlights once more the need for deep and large-scale imaging of galaxies in the 21-cm line of \ion{H}{i} in order to understand the role of interaction and feedback between galaxies and the IGM.
  
  \subsection{Gas accretion}
  
  Alternatively, some of the observed asymmetries in the \ion{H}{i} disc of NGC~300 could be the result of cold-mode accretion flows of primordial gas. Lopsidedness of the stellar and gaseous discs of galaxies is a well-known and common phenomenon \citep[e.g.,][]{Baldwin1980,Richter1994,Rix1995,Haynes1998}. Among other processes \citep{Jog2009}, cold-mode accretion flows of primordial gas have been discussed as the potential origin of the observed asymmetries \citep{Bournaud2005,Sancisi2008}.
  
  Hydrodynamical simulations by \citet{Keres2005} have shown that there are two distinct modes of gas accretion on to galaxies: hot-mode accretion of shock-heated gas at temperatures in excess of the virial temperature of the dark matter halo, $T > T_{\rm vir} \approx 10^{5}$ to $10^{7}~\mathrm{K}$, and cold-mode accretion of gas at lower temperatures of $T < T_{\rm vir}$. While in their simulations the cold mode dominates accretion at higher redshifts, the hot mode becomes increasingly more important at lower redshifts, as many dark matter haloes exceed a critical virial mass of about $10^{11.3 \ldots 12}~\mathrm{M}_{\odot}$ above which accreted gas becomes predominantly shock-heated \citep{Keres2005,Keres2009,Dekel2006}. Furthermore, cold-mode accretion is geometrically different from hot-mode accretion in that it occurs along filaments and therefore is not radially symmetric.
  
  Asymmetric cold-mode accretion could have resulted in asymmetries in the outer gas disc of NGC~300, because the accreted gas would end up in non-circular orbits with substantial orbital time-scales in excess of about $1~\mathrm{Ga}$ near the outer edge of the gas disc. This scenario is supported by numerical simulations of \citet{Bournaud2005} who conclude that accretion along filaments of the cosmic web with accretion rates of a few $\mathrm{M}_{\odot} \, \mathrm{a}^{-1}$ is likely to be responsible for the ubiquity of lopsidedness observed in the stellar and gaseous discs of galaxies.
  
  We conclude that sensitive \ion{H}{i} observations of a large sample of galaxies would provide the opportunity to systematically study the kinematics of gas in the outer disc to determine the contribution of external processes such as ram pressure or gas accretion.
  
  \section{Summary}
  \label{sect_summary}
  
  We have used the Australia Telescope Compact Array to map a large area of about $2^{\circ} \times 2^{\circ}$ (corresponding to a projected size of about $65 \times 65~\mathrm{kpc}^{2}$) around the Sculptor group galaxy NGC~300 in the 21-cm line of \ion{H}{i}. We achieved a $5 \sigma$ \ion{H}{i} column density sensitivity of $1.0 \times 10^{19}~\mathrm{cm}^{-2}$ over $8~\mathrm{km \, s}^{-1}$ for emission filling the synthesised beam of $180'' \times 88''$ FWHM. The corresponding $5 \sigma$ \ion{H}{i} mass sensitivity is $1.2 \times 10^{5}~\mathrm{M}_{\odot}$ under the assumption of a distance of $1.9~\mathrm{Mpc}$ for NGC~300.
  
  The major results and conclusions from our analysis of the \ion{H}{i} emission in NGC~300 are:
  
  \begin{itemize}
    \item NGC~300 is characterised by a dense inner disc with column densities in excess of about $5 \times 10^{20}~\mathrm{cm}^{-2}$ and an extended outer disc with column densities of typically less than $2 \times 10^{20}~\mathrm{cm}^{-2}$. The inner disc has an orientation angle of $290^{\circ}$ and is perfectly aligned with the optical (stellar) disc. The outer disc has a totally different orientation angle of $332^{\circ}$, and its major axis spans more than $1^{\circ}$ on the sky equivalent to a linear diameter of about $35~\mathrm{kpc}$ within the $10^{19}~\mathrm{cm}^{-2}$ column density contour.
    \item Under the assumption that the gas is optically thin we derive a total \ion{H}{i} mass for NGC~300 of $1.5 \times 10^{9}~\mathrm{M}_{\odot}$. This is only slightly short of the single-dish value of $1.7 \times 10^{9}~\mathrm{M}_{\odot}$ found by \citet{Koribalski2004}, indicating that the ATCA recovered about 90~per cent of the total flux. Of the total \ion{H}{i} mass of NGC~300, about $8.5 \times 10^{8}~\mathrm{M}_{\odot}$ (or 57~per cent) is contributed by the inner disc and $6.5 \times 10^{8}~\mathrm{M}_{\odot}$ (or 43~per cent) by the outer disc.
    \item We fitted a tilted ring model to the velocity field of NGC~300 to determine the \ion{H}{i} rotation curve out to a radius of $18.4~\mathrm{kpc}$. The derived rotation curve rises out to a radius of approximately $10~\mathrm{kpc}$ where we observe a maximum velocity of about $100~\mathrm{km \, s}^{-1}$. Further out the rotation curve slowly decreases to about $83~\mathrm{km \, s}^{-1}$ in the outermost ring at $18.4~\mathrm{kpc}$ radius.
    \item We constructed different mass models of NGC~300 involving a stellar disc, a gas disc, and a dark matter halo. We obtain good fits for both the Burkert and NFW dark matter halo models, both of which successfully cope with the decreasing rotation velocity in the outer parts of the galaxy. Fits with either Newtonian dynamics or Modified Newtonian Dynamics (MOND) and no dark matter halo at all result in much higher $\chi^{2}_{\rm red}$ values and unrealistic mass scaling factors for the gaseous disc.
    \item The twisted structure of the \ion{H}{i} disc of NGC~300 suggests distortion by tidal interaction, but no nearby companion galaxy is currently known that could have exerted strong tidal forces on the disc. At the same time, optical studies by \citet{Bland-Hawthorn2005} provide evidence against tidal interaction by demonstrating that the stellar disc of NGC~300 has a simple exponential radial profile out to the largest detected radii without any sign of distortion.
    \item Optical observations by \citet{Vlajic2009} have shown that the stellar population in the outer disc is predominantly old ($> 1~\mathrm{Ga}$) and that therefore the outer gas disc does not currently play a significant role in the star formation activity of NGC~300.
    \item Asymmetries in the outer \ion{H}{i} disc suggest that NGC~300 is affected by ram pressure interaction while moving through the intergalactic medium of the Sculptor group, with its proper motion on the sky directed approximately to the south-east. Estimates show that, under reasonable assumptions on the IGM density and relative velocity, the outer edge of the \ion{H}{i} disc of NGC~300 may be significantly distorted by ram pressure. From our analysis we conclude that the density of the IGM in the Sculptor group must be lower than about $10^{-5}~\mathrm{cm}^{-3}$ as otherwise the outer parts of the \ion{H}{i} disc of NGC~300 could not have survived.
  \end{itemize}
  
  \section*{Acknowledgments}
  
  We thank S.-H.~Oh for helpful discussions on the calculation of stellar mass-to-light ratios for the \textit{Spitzer} IRAC bands and W.~J.~G.~de~Blok for valuable comments that helped to improve the manuscript. The Australia Telescope Compact Array is part of the Australia Telescope which is funded by the Commonwealth of Australia for operation as a National Facility managed by CSIRO. This work is based in part on observations made with the Spitzer Space Telescope, which is operated by the Jet Propulsion Laboratory, California Institute of Technology under a contract with NASA.

  \bsp
  \label{lastpage}
\end{document}